# Chemically reversible isomerization of inorganic clusters

*Curtis B. Williamson[†§], Douglas R. Nevers[†§], Andrew Nelson[‡], Ido Hadar[∥], Uri Banin[∥]\*, Tobias Hanrath[†]\*, and Richard D. Robinson[‡]\**

[†]Robert F. Smith School of Chemical and Biomolecular Engineering, [‡]Department of Materials Science and Engineering, Cornell University, Ithaca, USA

[∥]Institute of Chemistry and the Center for Nanoscience and Nanotechnology, The Hebrew University, Jerusalem 91904, Israel

\*corresponding authors

Curtis B. Williamson
Cornell University
cw668@cornell.edu

Douglas R. Nevers
Cornell University
drn43@cornell.edu

Andrew Nelson
Cornell University
awn32@cornell.edu

Ido Hadar
The Hebrew University
ido.hadar@northwestern.edu

Uri Banin\*
The Hebrew University
uri.banin@mail.huji.ac.il

Tobias Hanrath\*
Cornell University
tobias.hanrath@cornell.edu

Richard D. Robinson\*
Cornell University
rdr82@cornell.edu

**Abstract**

Structural transformations in molecules and solids have generally been studied in isolation, while intermediate systems have eluded characterization. We show that a pair of CdS cluster isomers provides an advantageous experimental platform to study isomerization in well-defined atomically precise systems. The clusters coherently interconvert over an ~1 eV energy barrier with a 140 meV shift in their excitonic energy gaps. There is a diffusionless, displacive reconfiguration of the inorganic core (solid-solid transformation) with first order (isomerization-like) transformation kinetics. Driven by a distortion of the ligand binding motifs, the presence of hydroxyl species changes the surface energy via physisorption, which determines "phase" stability in this system. This reaction possesses essential characteristics of both solid-solid transformations and molecular isomerizations, and bridges these disparate length scales.

Phase transitions in solids and molecular isomerizations occupy different extremes for structural rearrangements of a set of atoms proceeding along mechanistic pathways. Phase transformations are initiated by nucleation events (*1*) that are difficult to define, and then propagate discontinuously from lattice defects with activated regions smaller than the crystalline grains (incoherent transformation) (*2*). Small molecule isomerization is a discrete process where the activation volume of the transition state is comparable to the size of the molecule (coherent transformation). Studies of isomerization and solid-solid transformations have thus far proceeded largely independently. Efforts to identify a system bridging these transformations have been made by examining the transformation of domains of reduced size, such as nanocrystals. Transformations of nanocrystals (100 to 10,000 atoms) do not mirror molecular isomerization, in that bulklike phase transition behavior extends to the nanometer length scale, even down to ~2 nm (*2*). Herein, we investigate the structural transformations in semiconductor cluster-molecules at the boundary between molecular isomerizations and solid-solid phase transitions in nanocrystals (**Fig. 1**) by studying magic size clusters (MSCs) (~10 to 100 atoms), as prototypical systems. Studies of these clusters (diameter <2 nm) with distinct chemical formulas revealed that the cluster structures were strongly influenced by the surface termination (*2–5*).

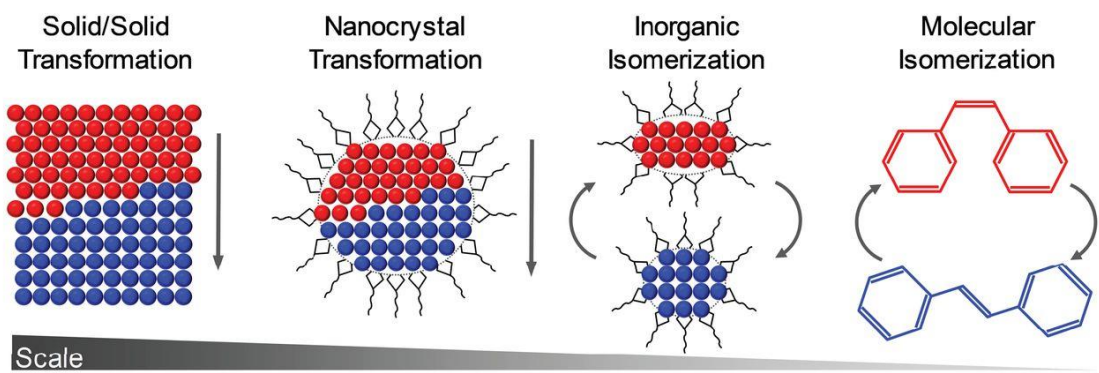

**Fig. 1: Inorganic isomerization**. Isomerization is well-established in small organic molecules (e.g., the cis-to-trans transformation of azobenzene), whereas bulk inorganic solids exhibit phase transformations. Although small in size, nanocrystals follow bulk-like behavior in their solid-solid transformations. At even smaller length scales, inorganic clusters isomerize with molecular- and inorganic-solid-like characteristics.

Previous work has observed that certain types, or families, of MSCs can be converted into other MSCs (*5–8*). Thus far, however, experiments claiming to have observed structural reorganization have been primarily conducted in the solution phase. Clusters in solution are free to interact with each other and with unbound surfactants, monomers, or byproducts, and these interactions promote mass transport and etching processes. For example, reports on InP clusters show irreversible structural changes, aggregation, and etching in the presence of high concentrations of amines (*5*). Such cases indicate a loss in the products' compositional integrity and thus that the transformation is not an isomerization. Structural transformations have been proposed for the same InP clusters at lower amine concentrations (*5*) and in CdS clusters following changes in temperature (*6*). In the former case, the assignment to a structural transformation was made by indirect methods (*9*) based on changes in $^{31}$P NMR shifts. This measurement permitted identification of only ~20% of the atoms in the cluster, none of which were directly associated with the surface ligands, and the experiment did not rule out the possibility of etching. Substantial changes in the $^{31}$P spectrum were observed in different solvents, bringing into question the dynamical stability of InP clusters in solution and, by extension, their status as isolated molecules undergoing discrete transformations. For the CdS clusters (*6*), the kinetics of incomplete transformations between cluster types indicated a very high activation energy (~3 eV), which is likely too large to account for merely structural reorganization energies and points instead to interparticle interactions. A primary complication of these solution-phase studies has been the lack of direct characterization of atomic structure, such as x-ray total scattering, in the native environment of transformation (*5, 6*) that can be used to identify the existence and extent of a structural transformation (*9*).

We demonstrate that a class of MSCs whose local structures can be modeled with a composition of $Cd_{37}S_{20}$ undergoes reversible isomerization between two discrete and stable states via a chemically-induced, diffusionless transformation. We preserved the composition by isolating our clusters in solid films and determined the cluster structures (fit residuals <0.2) through analysis of their x-ray pair distribution functions (PDF). Switching between the isomers was triggered by the absorption/desorption of water or alcohol (hydroxyl groups) with an activation barrier of ~1 eV in both directions. This chemically-induced reversible transformation has characteristics of both molecular isomerization and bulk solid-solid transformations. These clusters are an attractive starting point to merge the long- and short-length scale descriptions of such transformations as mediated by the external surface energy.

We synthesized high-purity clusters (*i.e.*, single product), characterized by a narrow excitonic absorption peak at 324 nm with negligible longer wavelength absorption, via our high concentration method (*10*). These clusters are stabilized by their mesophase (*11*) and immobilized in a thin solid film (**Fig. 2A**). We refer to this cluster type as α-$Cd_{37}S_{20}$.

After exposure of α-$Cd_{37}S_{20}$ films to methanol vapor, the exciton (first absorption) peak diminished and a second narrow absorption peak emerged at 313 nm, indicating formation of the new species, β-$Cd_{37}S_{20}$, with an energy gap larger by 140 meV. Hereafter, only transformations with methanol are discussed in detail, but any hydroxyl-bearing species (alcohol or water) can initiate conversion of α to β.

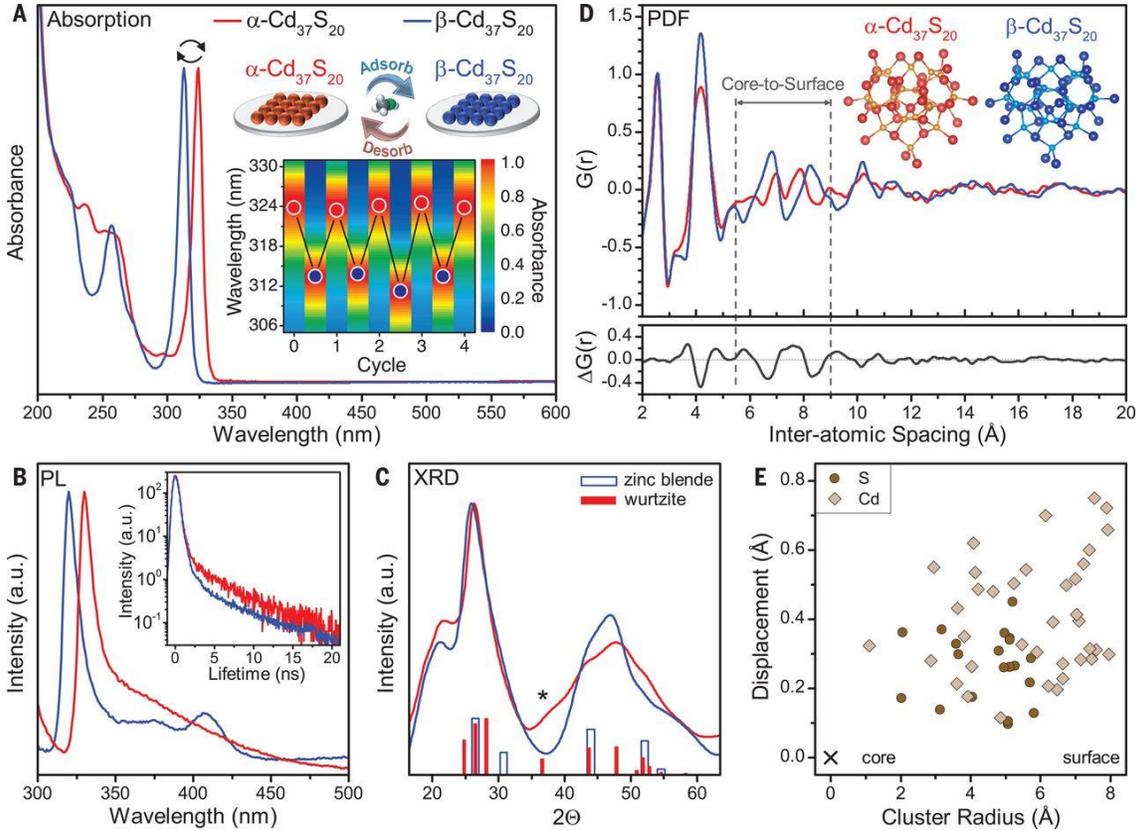

**Fig. 2: Electronic and structure analysis**. (**A**) Absorption spectra of pristine cluster isomers, α-$Cd_{37}S_{20}$ and β-$Cd_{37}S_{20}$, with excitonic peaks at 324 nm and 313 nm, respectively. The two isomers switch reversibly upon alcohol adsorption and desorption (inset schematic and contour plot). The slight deviation between cycles is associated with ambient temperature fluctuations. (**B**) Photoluminescence (PL) and lifetime (inset) of the isomers. (**C**) Synchrotron XRD patterns of α and β isomers referenced to a Cu-Kα source (λ = 0.154 nm). Peak positions for the wurtzite (with defining feature at 37°, asterisk) and zincblende phases of CdS are from Joint Committee on Powder Diffraction Standards card nos. 00-041-1049 and 00−010−0454, respectively. (**D**) Pair distribution functions (PDF) of the α and β isomers. Δ$G(r)$ = $G_α(r)$–$G_β(r)$ is the difference in the PDF between the two isomers and is largest for core-to-surface atom pair distances. Inset are the fitted structures of the α and β isomers with residuals of ~0.18. (**E**) Radial displacement of atoms between the α and β isomer structures with respect to distance from the cluster geometric center.

The β-$Cd_{37}S_{20}$ could be transformed back to α-$Cd_{37}S_{20}$ (reversion) by purging the methanol and heating the MSC film (> 60°C), and the reversion rate increased with temperature. We demonstrated a high degree of reversibility with four complete conversion-reversion cycles (**Fig. 2A,** inset) in the MSC isomerization without creation of other MSC families or nanocrystals (**Fig. S1D**). This behavior in MSCs is reminiscent of reversible isomerization reactions as are well-known in small molecules (*12*). We observed that differences in the dielectric environment only weakly alter the absorption maximum wavelength and that, in fact, the presence of hydroxyl-bearing species exclusively determines the favored isomer under the temperatures applied here. α-$Cd_{37}S_{20}$ may be stabilized at lower temperatures by maintaining an anhydrous environment, and β-$Cd_{37}S_{20}$ may be stabilized at higher temperatures (e.g., up to the boiling point of methanol) by maintaining the saturation of hydroxyl-bearing species. The stabilization of different MSC forms within a mesophase may have interesting consequences for nanoparticle formation once growth (e.g., by oriented attachment) is initiated, as mentioned in recent reports (*13*).

Both MSC isomers had low photoluminescence (PL) quantum yields (< 2.5%), which indicates that nonradiative decay processes dominated at room temperature. Substantial emission from the clusters electronic transitions was present (**Fig. 2B**). The PL decay transients could be fit by double exponentials (**Fig. 2B,** inset); the lifetimes of the slower decay rates are 5.8 and 5.2 ns for α-$Cd_{37}S_{20}$ and β-$Cd_{37}S_{20}$, respectively. The corresponding nonradiative and radiative rate constants for α-$Cd_{37}S_{20}$ are therefore $4.3 \times 10^6$ and $1.7 \times 10^8$ $s^{-1}$, respectively; for β-$Cd_{37}S_{20}$, they are $3.3 \times 10^6$ and $1.9 \times 10^8$ $s^{-1}$, respectively (see **SI** for calculations).

We analyzed the structure of the isomers by x-ray diffraction, noting that the XRD shoulder at 2Θ ~ 37° in the α-$Cd_{37}S_{20}$ is absent in the β-$Cd_{37}S_{20}$ (**Fig. 2C**). Although the peaks are broad, we interpret the α and β isomers as generally having "wurtzite-like" and "zinc blende-like" phases, respectively. We resolved the detailed atomic structure of the cluster isomers by fitting the pair distribution function (PDF) derived from the total scattering function (*G(r)*) using a Monte Carlo algorithm (see **SI Methods**). The best-fit structures of α and β (residuals of ~0.18 **Fig. S2B,C**) were comparable to InP clusters (*14*) (formula unit: $In_{37}P_{20}$), but with the substitution of In and P atoms for Cd and S, respectively. PDF analysis is an effective tool for atomic modeling that resolves fine features and subtle difference between data. Although powerful for low symmetry and disordered systems, atomic positions from PDF analysis and modeling hinges on the accuracy of the initial inputs (*15*, *16*). Repeated fitting showed that α- and β-$Cd_{37}S_{20}$ structures occupied unique energy minima whose separation greatly exceeded possible overlap from thermal displacements, so that the clusters' local structures are unambiguously distinct (**Fig. S2D**). Simulations including contributions from the organic ligands and the mesophase assembly determined that the organic ligand shell does not significantly contribute to scattering above Q = 1.5 Å$^{-1}$, where scattering from the inorganic structure is dominant, thus fitting *G(r)* beyond 2 Å even without organic contributions correctly resolves the positions of Cd and S (**Fig. S2B,C**). Our structures have a low symmetry (**Fig. 2D**, inset), unlike the highly symmetric tetrahedral coordination as reported for other CdS or CdSe MSCs (*17*, *16*). We hypothesize that our clusters resemble the InP structure because our clusters are similarly passivated with only carboxylate ligands, whereas the previously reported CdS or CdSe cluster structures are stabilized by amines, thiols, or a mixture of ligands. The representative structures of the clusters are molecular-like, but have scattering features similar to CdS crystals.

The difference between the α and β PDFs, $\Delta G(r)$, indicates changes in the atomic positions (**Fig. 2D**), where larger magnitudes signify a greater shift between the structures. Although $\Delta G(r)$ revealed preservation of the CdS bond lengths ($\Delta G(r) \simeq 0$ between 2.50-2.55 Å), there were appreciable differences in the bond angles ($\Delta G(r) \neq 0$ between 4-5 Å). Analysis of our atomic structures indicated an overall broader distribution of bond angles in the α-$Cd_{37}S_{20}$ than the β-$Cd_{37}S_{20}$ (**Fig. S2E,F**). These changes in conformation (atomic orbital overlap) must be the origin of the change in the excitonic gap between clusters. The greatest difference between the PDFs of the isomers was within the range of 5.5 to 9 Å, a range that corresponds to atomic pairs composed of one "core" atom and one "near-surface" atom.

Beyond an interatomic spacing of 12 Å, the $G(r)$ has oscillations that propagate to larger spacings (>30 Å). These features correspond to preferred intercluster orientations (diffraction texturing), which are broadened by variations in the cluster-cluster orientations (*18*). Our previous investigation revealed that these clusters form long-range assemblies (*11*). While texturing or preferred nanograin orientation can create challenges in structural analysis by x-rays, these challenges are less significant in PDF analysis (*18–20*). To assign a degree of transformation, we calculated the set of displacements required to transform one cluster into the other (**Fig. 2E**). The resulting relative displacements between isomers increases with radial distance from the cluster geometric center. Despite the large magnitude of displacement (up to ~30% of the Cd-S bond length for surface atoms), the connectivity of α- and β-$Cd_{37}S_{20}$ does not change. Therefore, the cluster isomerization is primarily displacive, characteristic of a solid-solid transformation, rather than reconstructive.

The FTIR spectra of α and β reveals that the isomerization stems from changes in the surface structure. We identify the carboxylate asymmetric stretches ($v_{as}$) at 1528 and 1538 cm$^{-1}$, respectively (**Fig. 3A,B**), and the carboxylate symmetric stretches ($\nu_s$) at 1410 cm$^{-1}$ for both isomers. The difference (Δ) between $\nu_{as}$ and $\nu_s$ gives the ligand binding motif: Δ<140 cm$^{-1}$ indicates a chelating bidentate configuration and Δ>140cm$^{-1}$ indicates a bridging bidentate configuration (**Fig. 3C**) (*21*). The dominant ligand configuration in the α and β isomers is the chelating bidentate configuration (Δ$_α$ = 118, Δ$_β$ = 128 cm$^{-1}$), but there is a strong shoulder in the $\nu_{as}$ (1580 cm$^{-1}$) in the α-Cd$_{37}$S$_{20}$ spectrum that points to the presence of some bridging ligands (Δ = 170 cm$^{-1}$) (*22*). Although this shoulder was absent in the β-Cd$_{37}$S$_{20}$, the spectral area of the $\nu_{as}$ between the isomers is preserved, implying no change in the overall ligand number. From the correlation of bond angles from single-crystal diffraction data to FTIR spectra for various metal carboxylates, (**Fig. S3B**) we estimated the change in the two sets of bond angles from Δ (*23*, *24*): the change in the chelating bond angle increased by ~0.5° upon conversion from α to β, and ligands changing from bridging to chelating configuration in the β-Cd$_{37}$S$_{20}$ decreased their bond angle by ~2.0°. The FTIR results indicated that the cluster isomerization is strongly coupled to a change in the ligand binding modes. We hypothesize that the modified ligand binding arrangement on the cluster surface is the chemical trigger to the isomerization.

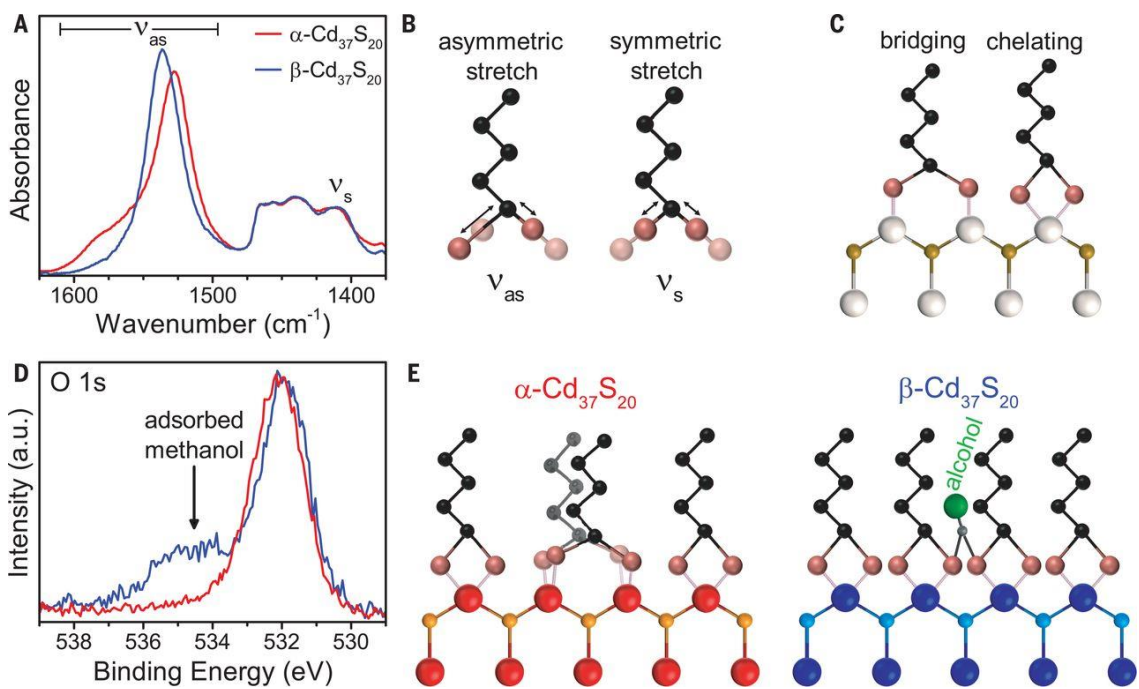

**Fig. 3: Organic Surface Analysis**. (**A**) FTIR spectra of the carboxyl asymmetric stretch ($v_{as}$) of the α-Cd$_{37}$S$_{20}$ and β-Cd$_{37}$S$_{20}$ isomers. (**B**) Schematic of the carboxylate stretch vibrations. (**C**) Observed bidentate carboxylate binding motifs. (**D**) O 1s XPS spectra in the α and β isomers. (**E**) Schematic of the ligand configuration on the isomer surface with chelating bidentate oleate molecules. Methanol hydrogen bonds with the oleate ligand to alter the chelating angle, which is larger in β-Cd$_{37}$S$_{20}$ relative to α-Cd$_{37}$S$_{20}$. Only one oleate is shown on each Cd atom for clarity.

X-ray photoelectron spectroscopy (XPS) showed that the Cd 3d spectra for α and β were not notably different (**Fig. S4A**, **Table S1, S2**), suggesting little interaction of the Cd atoms with adsorbed methanol. However, the O 1s spectrum for α-Cd$_{37}$S$_{20}$ showed a peak at 531.9 eV, which shifts to 531.7 eV in β-Cd$_{37}$S$_{20}$ spectrum. A second peak for β-Cd$_{37}$S$_{20}$ was present at 534.2 eV, which is attributed to physisorbed methanol (**Fig. 3D**). There was no evidence of dissociated methoxy species, which would have an O 1s peak at energies <532 eV (*25, 26*).

In combination, FTIR and XPS analyses indicate that the presence of methanol shifts the configuration of ligands bound to the surface of the cluster. Changes in the carboxylate angle result in a reconfiguration of Cd and S atoms at the cluster surface, initiating the overall isomerization of the cluster (**Fig. 3E**). Control experiments using aprotic solvents with strong to weak dielectric constants (acetone to perfluorohexane, respectively) (**Table S3**) did not induce a transformation (**Fig. S6A**). The β-$Cd_{37}S_{20}$ is formed following the adsorption of methanol on the surface of the cluster, which arises via hydrogen bonding with the oleate ligand (**Fig. 3D**). Hydrogen bonding, and not changes in the dielectric environment, distorts the carboxylate bond angle and initiates the necessary surface reconfiguration that induces the cluster isomerization. Interestingly, such a hydroxyl-triggered phase change in the similarly-structured $In_{37}P_{20}$ cluster (**Fig. S5**) was not spectroscopically observed (*14*). Why $In_{37}P_{20}$ lacked another stable polymorph under conditions similar to those applied here is not obvious. We suggest that further investigations should identify, with atomic precision, the differences in ligand conformation/binding and density, between $In_{37}P_{20}$ and $Cd_{37}S_{20}$.

The absorption peak of the β-$Cd_{37}S_{20}$ red-shifts to 320 nm (**Fig. S7A**) if methanol is not present (e.g., in vacuum), forming another species which we term β'-$Cd_{37}S_{20}$. Re-exposure to methanol rapidly regenerated β-$Cd_{37}S_{20}$. The details of the β and β' spectra were otherwise nearly identical, indicating that the β-like structure is metastable at low temperatures and that hydroxyl is only required as an initiator. The β-to-β' transition shows that absorbed methanol is not an essential contributor to the electronic structure. Likewise, there is no substantial differences between the β and β' XRD and PDF patterns (**Fig. S7D**), implying that the desorbed methanol affects the excitonic gap by way of dielectric effects.

Because the spectral overlap between the exciton of α- and β-$Cd_{37}S_{20}$ was small, we performed in-situ time-resolved spectroscopy measurements at temperatures 25-100°C to extract kinetic rate constants (**Fig. 4A**, **Fig. S8A** & **Table S4**) through the evolution of the first absorption peak of α-$Cd_{37}S_{20}$. The isomerization followed first-order reaction kinetics and had a small transformation hysteresis (**Fig. 4A**, inset). For α-to-β conversion, we kept the methanol partial pressure saturated; when the methanol partial pressure fell, the transformation deviated from first order. In a dry or high-temperature environment, the reverse transformation was also first order. The Arrhenius prefactor, $A$ (**Fig. 4B, Table S5**) was $3.4 \times 10^{12}$ s$^{-1}$, which corresponds to a vibrational frequency of a transformation across the transition state ($k_b T/h = 6.2 \times 10^{12}$ s$^{-1}$ at 300 K) and agrees with measured prefactors for adsorption/desorption and solid-solid transformation processes (*27, 28*). We observed a smaller reversion prefactor ($9.3 \times 10^9$ s$^{-1}$), on the order of those observed in some solid-solid transformations (*29*). Correspondence between the kinetic parameters from the optical experiments to those found from in situ diffraction confirmed the lack of structural intermediates (**Fig. S8F-H**), as did the isosbestic points in the optical absorption (**Fig. S1C**).

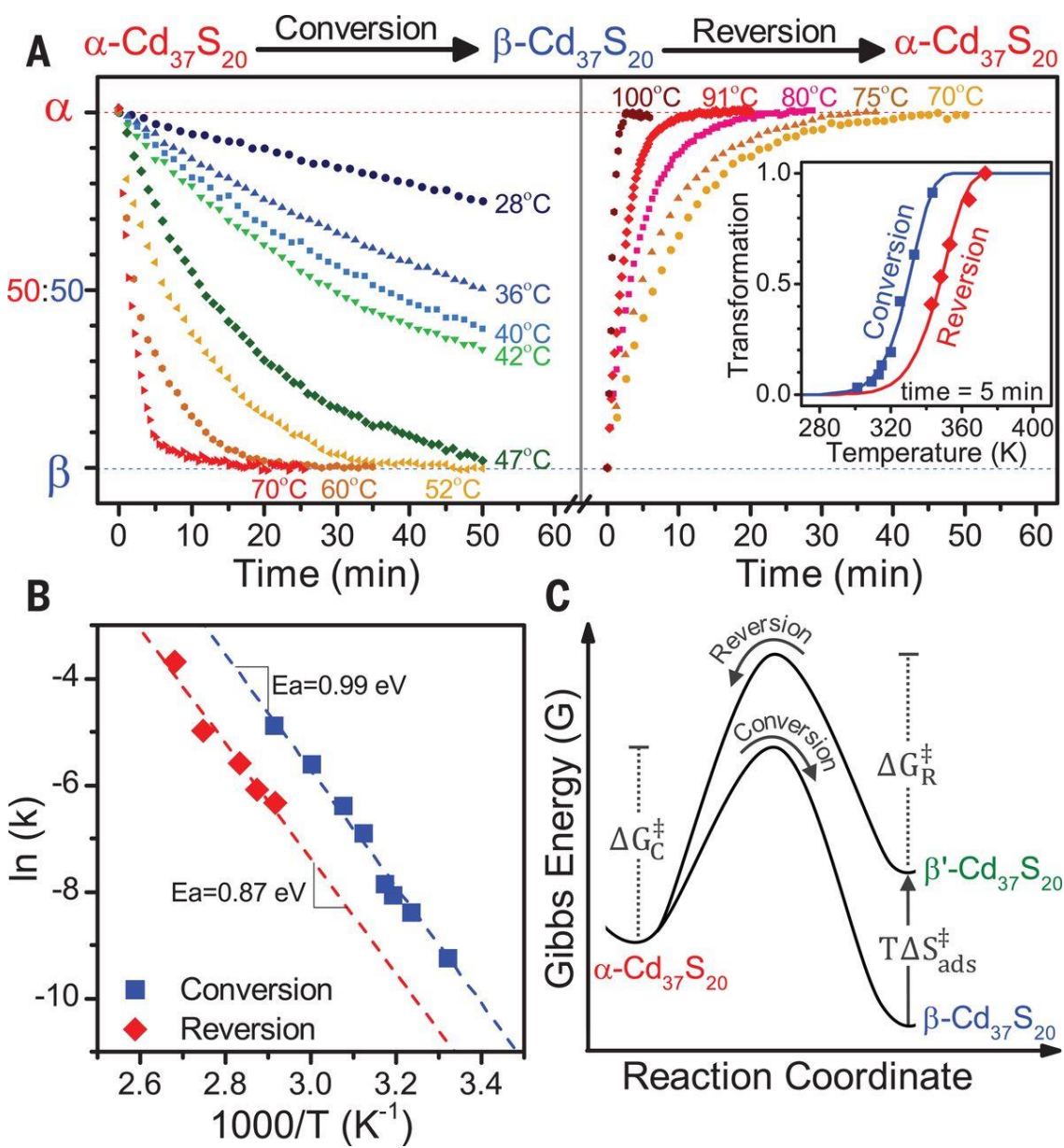

**Fig. 4: Transformation kinetics and thermodynamics.** (**A**) Kinetics of conversion and reversion processes. Both are first order: rate = $e^{-kt}$. Inset: hysteresis diagram for the transformed fraction at 5 min. reaction time. (**B**) Arrhenius plot for the transformation kinetics with fits (dashed lines). (**C**) Reaction coordinate diagram of the reversible transformation. The Gibbs free energies of the transition state for conversion and reversion, $\Delta G_C^\ddagger$ and $\Delta G_R^\ddagger$ respectively, are the same. $\beta$-Cd$_{37}$S$_{20}$ transforms to $\beta'$-Cd$_{37}$S$_{20}$ upon alcohol desorption with an entropic shift ($T\Delta S_{des}^\ddagger$).

The $E_a$ values for the conversion and reversion processes were 0.99±0.04 and 0.87±0.08 eV (95.5 and 84.0 kJ/mol), respectively. In comparison, first-principles calculations have shown that the binding energy of carboxylic acids onto ($Cd_{33}Se_{33}$) is ~0.7 to 1.5 eV, with larger values for binding on higher index facets (*30*). Compared to previously reported energies for a similarly-described MSC partial transformation performed on unpurified samples in dilute solution, our activation energies are a factor of three smaller and align more closely with common structural transformation energies (i.e., solid-solid transformation and isomerization) (*6*). Our lower activation energy from more rigorous experiments better agrees with the low degree of local structural change during the conversion as inferred from direct characterization methods, such as pair distribution analysis.

We used the Eyring equation to derive the Gibbs free energy of the transition state, $\Delta G^\ddagger$, (**Table S6**) and the apparent values for the enthalpy and entropy of the transition state ($\Delta H^\ddagger$ and $\Delta S^\ddagger$, respectively) (**Fig. 4C**). The $\Delta H^\ddagger$ for the conversion and reversion processes are 0.96±0.04 and 0.84±0.07 eV, respectively. The difference in $\Delta H^\ddagger$ between the processes may be related to the non-equilibrium desorption of physisorbed methanol in the reversion process. To investigate the possibility of chemisorption and steric interactions, we performed the reversion process on $\beta$-$Cd_{37}S_{20}$ produced from alcohols with increasing alkyl chain length (**Fig. S8H**) and found that $\Delta G^\ddagger$ was independent of the alcohol. We conclude the $\Delta H^\ddagger$ is predominantly the free energy to relax the inorganic core following the change in the boundary conditions of the chemical potential at the ligand-core interface. Because we were unable to isolate the two isomers in coexistence with each other, the transformation must be kinetically controlled and comparison to thermodynamic parameters, such as enthalpies of mixing (which are 1000-fold smaller (*31*)) cannot be made. The weak temperature dependence implies that the transformation is predominantly enthalpic, and the mean difference in $\Delta S^{\ddagger}$ of the transformation ($\Delta S^{\ddagger}_{conversion}$ - $\Delta S^{\ddagger}_{reversion}$) of +0.52 meV/K is consistent with H-bonding entropies. As implied by **Fig. 4C**, the α-$Cd_{37}S_{20}$ structure becomes thermodynamically unstable with respect to thermal decomposition into β-$Cd_{37}S_{20}$ due to changes in the surface energy. Removal of the surface energy perturbation by desorption of hydroxyl raises the free energy of β-$Cd_{37}S_{20}$, likewise rendering it thermally unstable to decomposition to the α form.

A coherent transition between two clusters implies conservation of binding coordination, a single rate constant for the reaction, and simultaneous transformation of the entire cluster rather than growth from a nucleation site (*2*, *32*, *33*). Based on the transformation kinetics and the lack of any observable intermediates, the upper bound on the lifetime of an intermediate state must be on the order of $10^{-13}$ s (see **SI** for calculations), a time scale comparable to bond vibrations (33) and the lifetime of molecular transition-states (*34*); additionally, to achieve the same rate of transformation for the MSC in an incoherent process, a phase boundary would need to move at a velocity comparable to the speed of sound of the bulk material (*28*, *35*). The small atomic displacements shown from PDF analysis indicate a structural reconfiguration without a change in coordination number. Our experimental kinetics are thus consistent with a coherent atomic displacement occurring in a single step across the entire cluster.

The authors thank Peter Ko and Michael Steigerwald. **Funding**: This work was supported in part by the National Science Foundation (NSF) under award Nos. CMMI-1344562 and CHE-1507753. U.B. acknowledges funding from the European Research Council (ERC) under the European Union's Horizon 2020 research and innovation programme (grant no. 741767). U.B. also thanks the Alfred & Erica Larisch memorial chair. This work also made use of the Cornell Center for Materials Research Shared Facilities, which are supported through the NSF MRSEC (Materials Research Science and Engineering Centers) program (Grant DMR-1719875). This work includes research conducted at the Cornell High Energy Synchrotron Source (CHESS), which is supported by the National Science Foundation and the National Institutes of Health/National Institute of General Medical Sciences under NSF award DMR-1332208. R.D.R. thanks the U.S. Fulbright Scholar and Hebrew University for partial funding during this work. **Author contributions**: [§] C.W. and D.N contributed equally. C.W. synthesized and prepared samples for UV-Vis absorption spectroscopy, FTIR spectroscopy, XPS spectroscopy, and kinetic analysis. C.W. and D.N. prepared samples for x-ray total scattering and diffraction. C.W., D.N., and A.N. contributed to the PDF analysis of the total scattering data. I.H. and U.B. acquired and analyzed fluorescence spectroscopy and lifetime measurements. All authors contributed to the interpretation of results and preparation of manuscript. **Competing interests**: none declared. **Data and materials availability**: all data needed to evaluate the conclusions in the paper are present in the paper or the Supplementary Materials.

**Supplementary Materials**

Materials and Methods

Figs. S1 to S10

Tables S1 to S6

References (36-48)

# Supplementary Materials for

## Chemically reversible isomerization of inorganic clusters

Curtis B. Williamson[§], Douglas R. Nevers[§], Andrew Nelson, Ido Hadar, Uri Banin*, Tobias Hanrath*, Richard D. Robinson*

[§]These authors contributed equally to this work.
*Corresponding authors: uri.banin@mail.huji.ac.il (U.B.); tobias.hanrath@cornell.edu (T. H.); rdr82@cornell.edu (R. R.)

**This PDF file includes:**

Materials and Methods
SupplementaryText
Figs. S1 to S10
Tables S1 to S6
References

**Materials**

The following chemicals were used as received. Oleic acid (OA, 90%), cadmium oxide (99.5%), ethyl acetate (≥99.5%), tri-n-octylphsosphine (TOP, 97%), sulfur (purified by sublimation, particle size ~100 mesh), hexane (95%, anhydrous), 1-propanol (≥99.5%, ACS reagent), 1-octanol (99%), and 2,2,2-trifluoroethanol (>99%, ReagentPlus) were purchased from Sigma-Aldrich. Methanol (99.8%, Certified ACS) was purchased from Fisher Scientific. Ethanol (200 proof, Anhydrous KOPTEC USP) was from Decon Labs. Perfluorohexane (99%) was from Alfa Aesar. Tetrahydrofuran (>99.9%, DriSolve, BHT stabilized) and acetone (99.5%, ACS grade) were from EMD Millipore.

**Methods**

*UV-Vis Absorption Spectroscopy* – Measurements of absorption spectra were performed from 200 to 800 nm on an Ocean Optics USB2000+ photodiode spectrometer with a DH-2000-BAL light source. All kinetic and in-situ absorption spectra were averaged over 25 scans with an integration time of 200 ms (5 s total), unless otherwise specified. Background subtractions were done with an empty FUV quartz cuvette.

*Cryogenic UV-Vis Absorption Spectroscopy* – Measurements were performed on a Cary-5000 spectrometer at a resolution of 0.1 nm using a Janis STVP-100 Optical Cryostat from 5 to 300 K (ambient). The clusters were measured as a thin film on a $CaF_2$ window. During measurement, temperature was within ±0.5 K of the set point. Data at room temperature were collected before cooling. After cooling, scans were collected at increasing temperature at 5, 10, and 20 K, then at 20 K intervals up to ~300 K. Background subtractions were done with a clean $CaF_2$ window at room temperature.

*Fourier-Transform Infrared Spectroscopy* – Measurements were performed on a Bruker Tensor II spectrometer connected to a Hyperion FTIR microscope with a MIR source and KBr broadband beamsplitter. All spectra were collected in a vapor cell with $CaF_2$ windows and a 10 kHz scanner velocity. Each spectrum is an average of 10 scans. Background subtractions were done with the blank vapor cell. All spectra were collected between 800-4000 $cm^{-1}$ with a resolution of 1 $cm^{-1}$.

*X-ray Scattering* – Total scattering measurements were performed at the F2 beamline at the Cornell High Energy Synchrotron Source (CHESS) (wavelength=0.20218 Å, Energy=61.32 keV, bandwidth=0.25%). Images were collected using a GE Flat Panel detector with a pixel size of 200 x 200 µm and total area of 2048 by 2048 pixels. The sample to detector distance was 238.78 mm, as determined from a $CeO_2$ standard. The patterns were averaged over 10 scans with 16 s collection time for each frame. Background subtraction was done with dark (no beam) and empty (beam present, but no sample) images. The averaged images were then integrated using Fit2D(*36*) and normalized to the incident beam intensity. Samples were analyzed as solids supported by steel washers.

*X-ray Photoemission Spectroscopy* – Measurements were performed on a Surface Science Instruments SSX-100 ESCA spectrometer with the following parameters: monochromated aluminum Kα source (1486.6 eV), beam spot size 1 mm, analyzer pass energy 50 V, angle between the electron energy analyzer and sample normal 55°, and operating pressure lower than $2x10^{-9}$ Torr.

*X-ray Diffraction* – XRD data were collected on a Bruker D8 Discover microbeam diffractometer with the following parameters: Cu Kα source (1.54 Å), 2 mm polycapillary collimator, and Vantec-500 area detector. All samples were washed with ethyl acetate and dried in vacuum prior to characterization. The powders were analyzed on $CaF_2$ window, and spots from the $CaF_2$ were masked before integration of the 2D X-ray image. The integrated data were background subtracted with a single decaying exponential and normalized to the peak at ~26° 2θ.

*Fluorescence Spectroscopy* – Photoluminescence was measured using an Edinburgh Instruments FL920 fluorescence spectrometer in L configuration. The spectrophotometer is equipped with steady state and ps pulsed sources (Laser and LED), excitation and emission are spectrally filtered through double monochromator to improve signal to noise. Emission is collected by a fast PMT in a single photon counting scheme (SPC). For steady state PL the sample is excited by the UV emission of a xenon lamp and the emission was scanned. Quantum yield (QY) was measured in a relative method - the emission intensity of the sample was measured and compared to emission intensity of a known fluorophore (naphthalene), with the same absorption.

*Fluorescence Lifetime* – Lifetime measurements were performed in time-correlated single-photon counting (TCSPC) approach using the spectrometer above with a TCC900 TCSPC card. A cuvette containing each of the solutions was excited at a wavelength of 270 nm by a picosecond pulsed UV light emitting diode (EPLED-270 Edinburgh Instruments), with pulse width of 700 ps and repetition rate of 1 MHz. The excitation power was attenuated by a variable neutral density filter. The emission from the sample was collected at a right angle, transferred through double monochromators to suppress the fundamental excitation light of the laser, and collected using a Hamamatsu R2658P PMT. The instrumental resolution function (IRF) of our spectrophotometer results is ~200 ps, with a FWHM ~0.9 nsec. To define the start of the emission beyond the IRF we fit the data starting around 2.8 ns. To normalize the data we used a value found from averaging in a window of ~2.7 – 3.1 ns. An average background was computed by averaging at long times; this was then subtracted from the data. The best fit to the data was with standard biexponential decay functions, the shorter time constant is faster than the instrumental resolution (~1.0 ns).

*Preparation of 1.0M Cadmium Oleate (CdOl)* – In a 50 mL round bottom flask (RBF) connected to a Schlenk line, 1.28 g (10 mmol) of CdO and 10 mL (8.95 g) of oleic acid are added. The contents are heated to 50°C and mixed with a 2 cm stir bar at 1000 rpm. At 50°C, the suspended mixture is placed under vacuum and degassed. When bubbling of the fluid stops, the RBF is placed under $N_2$ gas and the flask is heated to 140°C. The CdO takes roughly 1-2 hours, at 140°C, to react completely with the oleic acid and makes a translucent and viscous tan-orange solution. Once fully reacted, the solution is cooled to 90°C. At 90°C, the mixture is placed under vacuum to remove the water produced from the reaction. *Note: This step is very sensitive to the vapor space and temperature. Below 90°C, the mixture is too viscous for bubbles to break and above 100°C, the vapor pressure is too high and the solution bumps.* While under vacuum and bubbling is under control, the solution is heated to 120°C. When bubbling subsides, the flask is cooled to 50°C and placed under $N_2$ gas.

*Preparation of 2.5M Tri-n-octyl Phosphine Sulfur (TOPS)* – In a 20 mL scintillation vial in a glove box, 0.40 g (12.5 mmol) of elemental sulfur and 5.0 mL (4.15 g) of tri-*n*-octyl phosphine are added. The contents are mixed with a 1 cm stir bar at 1000 rpm. *Caution: Larger quantities (>20 mL) of TOPS produce excessive heat and may require* *cooling or slower addition of tri-n-octyl phosphine to the sulfur.* Once fully dissolved, the vial can be brought out of the glove box and is air stable.

*F324 Magic-sized Cluster Reaction* – While the CdOl solution is at 50°C, 2.0 mL of the 2.5 M TOPS solution is injected into the 10 mL of 1.0M CdOl solution. The solution is mixed for 5 min to ensure a homogenous concentration. **Comment:** *At this point, the mixture can be stored in air by cooling to room temperature and used at any time.* The solution is then heated to 140°C over a 20 min period. Once the temperature reaches 140°C, the mixture reacts for 120 min. After which, the reaction is cooled to 100°C and then quenched with 10 mL of ethyl acetate, which produces a white precipitate.

*Conversion Process* – The conversion process uses thin films drop cast from a solution of F324 dissolved in hexane. A heating stage is heated to a pre-determined temperature and allowed 60 min to equilibrate before inserting the F324 thin film. Once the temperature has re-equilibrated following insertion, methanol is injected directly to the bottom of the cuvette (away from the thin film), which vaporizes and forms a saturated vapor phase. If liquid methanol comes in contact with the film or condenses on the thin film, the films rapidly become turbid and initiates mass transport processes. Precautions were taken to measure and account for partial pressures. The methanol vapor initiates the transformation of the thin film, and the absorption spectra are collected at once and every 5 s thereafter. See **Calculations** for interpretation of conversion kinetics. **Note:** *It is crucial that liquid methanol does not form on the film as this promotes nanoparticle growth.*

*Reversion Process* – The conversion process uses thin films drop cast from a solution of F313 dissolved in hexane. The heating stage is heated to a pre-determined temperature and allowed 60 min to equilibrate before inserting the F313 thin film. The absorption spectrum is recorded immediately and every 5 s thereafter. There is rapid peak shifting due to heating (see **Fig. S6** for exciton-temperature relation), which decreases substantially after 30 s. Once exciton shifting subsides, further changes in the absorption spectrum are attributable to the reversion process. See **Calculations** for interpretation of reversion kinetics.

*Transformation Cycles* – Cycling of the cluster film uses the same protocol for the F324 film preparation and the transformation protocol for the conversion and reversion processes. For the conversion component of the cycling, the F324 film is heated to 59°C and exposed to methanol. After 60 min, the film is cooled to room temperature and the absorption spectrum of the F324 is recorded. Methanol is removed by purging with $N_2$ gas, then the spectrum of F320 is recorded. For the reversion component of the cycle, the F313 film is heated to 80°C. After 60 min, the thin film is cooled to room temperature. At room temperature, the film is flushed with $N_2$ and the absorption spectrum of the F324 is recorded. The conversion and reversion processes described in this section are repeated three more times (a total of 4 cycles). The slight shifts in the absorption peak positions are thermal shifts due to changes in the ambient temperature of the laboratory (see **Fig. S6** for exciton-temperature relation).

*Exciton-Temperature Relation* – To investigate the effects of temperature on the clusters, F324 and F313 were made into thin films on various substrates with optical densities of the excitonic peak within the range of 0.4-0.8. For temperatures above ambient (>300 K), the cluster films were prepared on fused silica cover slips and placed onto a heating stage. The cluster films were exposed to ambient air throughout the heating experiment. The thermocouple was placed directly onto the cluster film immediately adjacent to the optical path of the spectrometer. See **Fig. S6** for the exciton-temperature relation for temperatures above ambient. For temperatures below ambient (<300 K), the cluster films were prepared on CaF$_2$ windows (similar to Preparation of F324/F313 films) and inserted into a cryostat. The film is placed under vacuum and then filled with helium. The absorption spectra were collected using a Cary 5000 spectrometer. The cluster films were cooled with helium to 5 K and absorption spectra were recorded every 20 K with an equilibration time of 5 min at temperature. ***Note: The vacuum/helium atmosphere in the cryostat causes the F313 to transition to the F320.*** See **Fig. S6** for the exciton-temperature relation for temperatures below ambient.

*Monte Carlo and PDF Analysis* – The integrated total scattering data is converted to the pair distribution function (PDF) to better resolve differences between the two physical structures. The PDF is calculated by following the work and calculations of Billinge et al(*37–39*), which is briefly described below. There available software package, PDFgetx3, converts the experimental integrated scattering data into PDF. However, to give us more flexibility with the PDF analysis and the ability to perform reverse Monte-Carlo, we script our own code (in MATLAB R2018a) using the work and calculations of Billinge et al. We verify the accuracy of our code by comparing our simulated and experimental PDFs to those produced by PDFgetx3.

Here, we describe briefly the steps and calculations used in our PDF analysis. More detailed descriptions and information of the steps can be found in the work of Billinge et al (*37*). The experimental integrated x-ray scattering data, $I(Q)$, possesses scattering intensities from all sources (e.g., elastic, inelastic, fluorescence). To convert the scattering data into the PDF, we first need to convert the $I(Q)$ into the structure function, $S(Q)$, and eliminate all other contributions (background) to the scattering intensity that is not the coherent elastic scattering of our sample. The scattering intensity, $I(Q)$, may be converted into the $S(Q)$, which is a normalization of the scattering intensity with respect to the average over all atomic scattering factors, $f(Q)$, in the sample.

$$S(Q) - 1 = \frac{I(Q)}{\langle f(Q) \rangle^2} - \frac{\langle f(Q)^2 \rangle}{\langle f(Q) \rangle^2} \qquad [\text{i}]$$

The structure function has an additive correction function applied to the function. Billinge et al. proposed that that the experimental structure function, $S_m(Q)$, deviates by a slowly changing additive factor, $\beta_s(Q)$, from the correct $S(Q)$ as:

$$S_m(Q) = S(Q) - 1 + \beta_s(Q) \qquad [\text{ii}]$$

The modified structure function, $S_m(Q)$, is then scaled by $Q$, to produce $F_m(Q)$, which oscillates around and approaches zero with increasing $Q$.

$$F_m(Q) = Q[S(Q) - 1] + Q\beta_s(Q) \qquad [\text{iii}]$$

To remove the correction factor, $\beta_s(Q)$, a polynomial function, $P_n(Q)$, is subtracted from $F_m(Q)$.

$$F_c(Q) = F_m(Q) - QP_n(Q) \qquad [\text{iv}]$$

Where the subscript, $n$, is the order of the polynomial fit and is determined by how far the scattering data extends into Q-space ($Q_{max}$), and by the value below the shortest bond lengths in the material ($r_{poly}$), as:

$$n = \frac{r_{poly} Q_{max}}{\pi} \qquad [\text{v}]$$

Given that the order of the polynomial fit, $n$, in most cases is not an integer, we took the average of the two different ordered polynomial functions, $P_n(Q)$, with $n$ being round to its two nearest integers. This average polynomial function is used to produce the $F_c(Q)$ that is then converted to the PDF, G(r).

$$G(r) = \frac{2}{\pi} \int_{Q_{min}}^{Q_{max}} F_c(Q) \sin Qr \, dQ \qquad \text{[vi]}$$

We use the following parameters in our PDF analysis: $Q_{min} = 1$ Å$^{-1}$, $Q_{max} = 16$ Å$^{-1}$, $r_{poly} = 0.9$ Å, $r_{min} = 0$ Å, $r_{max} = 40$ Å. $Q_{min}$ and $Q_{max}$ are the limits in the range of scattering angle included in the Fourier transform to obtain the PDF. The value of $r_{poly}$ is commonly selected for analysis and is below all bond lengths that would exist in our material. $r_{min}$ and $r_{max}$ are the range of the calculated PDF.

To determine the 3-D structure of our clusters, we first simulated various x-ray diffraction patterns using the Debye scattering equation. Again, we script our own code, using MATLAB R2018a, following the works and calculations of Billinge et al (*39*). The scattering intensity profile (XRD) of our structures are calculated from:

$$I(Q) = \sum_i \sum_j f_i f_j \frac{\sin Q r_{ij}}{Q r_{ij}} \qquad \text{[vii]}$$

Where $f_{i,j}$ is the atomic scattering factor of atoms $i,j$ and $r_{ij}$ is the interatomic distance between atoms $i$ and $j$. The atomic scattering factors are retrieved from an international database table (*40*). The simulated scattering intensity is then converted into the PDF following the same protocol as the experimental data, starting with eq. [i].

We explored a plethora of structures from literature and our own development. The structure with the closest resemblance to our experiment integrated intensity data is the atomic structure of an InP magic-sized cluster determined from single crystal, by the work of Cossairt et al (*14*). We focus our reverse Monte Carlo and structural analysis around this InP structure to find the structure of our cluster isomer. For our analysis, we replaced the In atoms with Cd atoms and the P atoms with S atoms. We have provided some analysis into a similarly sized CdS tetrahedral cluster, as this structure was the expected structure for CdS MSCs by previous reports in the literature (*16, 17*). The reverse Monte Carlo (reverse MC) we employ is an iterative process in which atoms are moved around the structure. Each time an atom is moved, the scattering intensity profile, $I(Q)$ is computed and then converted to $G(r)$ to compare against the experimental data. The simulated $G_{calc}(r_i)$ is compared to the experimental $G_{obs}(r_i)$ through the residuals function (*38*):

$$R_w = \sqrt{\frac{\sum_{i=1}^{N}[G_{obs}(r_i) - G_{calc}(r_i)]^2}{\sum_{i=1}^{N} G_{obs}^2(r_i)}} \qquad \text{[viii]}$$

We identify completion of the reverse MC when the accepted moves become sparse and the residual appears to have plateaued. An accepted move must meet the following criteria: 1) the atom being moved does not become unphysical, and 2) the calculated residual of the move is less than the previous accepted move. These two simple criteria ensure stability in the algorithm and mimic phenomenon similar to a solid-solid transformation. An unphysical move is defined as 1) two or more atoms that are separated by distances less than an expected bond length for Cd-S (2.54±0.19 Å), and 2) atoms having zero bonds (i.e., atoms on the surface are drifting away from the cluster). General guidelines by a leading group in the PDF community (Simon Billinge et al.) uses a residual value of 0.2 as an upper-boundary; below this value the fits and atomic models are accurate (*16, 41*). For our candidate structures, the residuals from our reverse MC have plateaued below this 0.2 threshold, which indicates that our structures are accurate. We perform statistics on our structure by running the algorithm multiple times for each cluster (see **Fig. S2F**).

The organics of our cluster isomers play a large role in the transformation. However, the scattering intensity of organic materials is significantly less than that of inorganic materials. We have identified a feature that is affiliated with the organic material in our clusters. This feature is at ~1.4 Å$^{-1}$ and relates to the oleate ligand. Unfortunately, it is not intuitive in how the ligands are structured in our material. Our approach to factor in the organic contribution to our cluster structure is to simulate the structure (using eq. [vii]) of an oleate single crystal (*42*) and then "melt" the structure through thermal broadening (see **Fig. S2M,N**). The number of oleates are proportional to the number of Cd atoms in our cluster. This 1:1 composition of oleates to Cd was determined in a previous report, in which the composition of the clusters is Cd$_2$S(Oleate)$_2$(*4*). A thermal broadening, Debye-Waller factor, or B-factor can be applied to the Debye scattering equation, which artificially broadens the peaks in the simulated diffraction patterns, effectively "melting" the structure. The simulated organic $I(Q)$ is added to the cluster $I(Q)$ and then converted to PDF.

The synthesis of these clusters occurs in tandem with the formation of an extended mesophase the dominates at small scattering angles (low Q, <1.0 Å$^{-1}$). This mesophase assembly causes slight texturing in the scattering pattern of the clusters depending on the sample preparation (pressed/shear) and the scattering techniques (reflection vs transmission) (see **Fig. S2J** for details). To account for this texturing, we simulated our mesophase by simulating multiple clusters within registry of one another, which narrows the peak at ~1.85, and then applied the Debye-Waller factor to vary cluster-cluster orientation. Then, we removed the single cluster scattering intensity. The remaining scattering intensity is only that of the mesophase assembly, in which the scattering intensity oscillates about zero. The sum of the scattering intensity for the mesophase is zero and this is because we are not introducing new scattering species. But introducing an interference effect as a result of cluster-cluster x-ray interactions. This simulated mesophase $I(Q)$ is added to the simulated cluster and organic $I(Q)$ to produce the pattern that appears in **Fig. S2B,C**.

**Supplementary Text**

Notation

The identification of the Cadmium Sulfide Magic-sized Clusters (MSCs) within the supporting information have been labeled differently from the main text. The labeling of the MSCs in the supporting information follows typical semiconductor MSCs identification: position of the 1$^{st}$ absorption peak at room temperature. Therefore, the α-$Cd_{37}S_{20}$ isomer is a 324 nm MSC family (F324); the β-$Cd_{37}S_{20}$ isomer is a 313 nm MSC family (F313); and the β'-$Cd_{37}S_{20}$ isomer is a 320 nm MSC family (F320).

α-$Cd_{37}S_{20}$ isomer = F324
β-$Cd_{37}S_{20}$ isomer = F313
β'-$Cd_{37}S_{20}$ isomer = F320

Calculations

*Gibbs Free Energy* – The Gibbs free energy, $\Delta G^{\ddagger}$, of the transition state is defined as

$$\Delta G^{\ddagger} = \Delta H^{\ddagger} - T\Delta S^{\ddagger} \qquad [1a]$$

where $\Delta H^{\ddagger}$ is the enthalpy and $\Delta S^{\ddagger}$ is the entropy of the transition state. We evaluate the value of $\Delta G^{\ddagger}$ at each temperature for the conversion and reversion processes to determine an average $\Delta G^{\ddagger}$. Since the entropy of both processes are small relative to the enthalpy, the value of $\Delta G^{\ddagger}$ is nearly constant. Below is a sample calculation for the reversion process at 70°C (343 K) and 100°C (373 K). Between the two end points of the reversion process, there is a small difference in the $\Delta G^{\ddagger}$. The difference is even smaller in the conversion process, since the measured $\Delta S^{\ddagger}$ is nearly zero.

$$\Delta G^{\ddagger} = 0.84 \, eV - 343 \, K \times (-0.67x10^{-3}\frac{eV}{K}) = 1.07 \, eV$$

$$\Delta G^{\ddagger} = 0.84 \, eV - 373 \, K \times (-0.67x10^{-3}\frac{eV}{K}) = 1.09 \, eV$$

*Interpretation of Transformation Kinetics* – The rates of transformation are determined from the evolution of the F324 optical density. For each conversion reaction, at *t=0* min, the absorption spectrum is that of a pure F324 with the optical density of the excitonic peak at 0.50±0.10. Likewise, for each reversion reaction, at *t=0* min, the absorption spectra is that of the pure F320 with the optical density of the excitonic peak at 0.50±0.10. With optical densities for the transformation processes being <1.0, the Beer-Lambert law posits a linear relationship of optical density to cluster concentration. This correlation enables normalization of the kinetic data between zero and one, such that the normalized optical spectrum is equal to the normalized concentration. Since the reaction goes to completion, the evolution of the absorption spectra can be fit with a linear combination of F324 and F313 for the conversion, where the sum of the fraction is ~1.0 at any time (see **Fig. S8B**). Similarly, the rate of depletion in the normalized exciton peak of the F324 matches that of the spectral fraction of the F324 pure component. The linear combination of pure components still holds true for the reversion reaction, so the absorption spectra can be fit with a linear combination of F324 and F320, where the sum of the fraction is ~1.0 at any time. The data for **Figure 4A** is the spectral fraction, $f_{324}(t)$, of the F324. We define the rate of conversion, $R_{conv}$, and reversion, $R_{rev}$, as;

$$f_{324}(t) + f_{313}(t) = 1 \qquad [2a]$$
$$R_{conv} = f_{324}(t) = e^{-kt} \qquad [2b]$$

$$R_{rev} = f_{324}(t) = 1 - e^{-kt} \qquad [2c]$$

where, $k$ is the rate constant. For the conversion reaction, there is full separation of excitonic peaks of the F324 and F313 for the conversion reaction. The analysis can be simplified to the only the decay rate of the F324 peak, since there is no optical contribution of the F313 spectra to the F324 excitonic peak position:

$$f_{324}(t) = \frac{F324(t)}{F324_0} = e^{-kt} \qquad [2d]$$

where $F324_0$ is the initial optical density of the F324 excitonic peak and $F324(t)$ is the optical density at time $t$.

*Radiative and Non-radiative Recombination* – Estimation of the MSC radiative and non-radiative recombination rates can be calculated from their relationship to the MSC quantum yield (QY). The first relationship, eq. [3a], predicts the radiative recombination rate and requires knowledge of the excited state lifetime:

$$QY = \gamma_{rad}\tau_{eff} \qquad [3a]$$

where $\gamma_{rad}$ is the radiative recombination rate and $\tau_{eff}$ is the effective lifetime. The second relationship, eq. [3b], states the quantum yield is the ratio of radiative recombination pathway to all other recombination pathways:

$$QY = \frac{\gamma_{rad}}{\gamma_{rad} + \gamma_{nonrad}} \qquad [3b]$$

where $\gamma_{nonrad}$ is the non-radiative recombination rate. The sample calculations below are for the determination of the F313 radiative and non-radiative recombination rates. These calculations are the same for the F324.

$$\gamma_{rad} = \frac{QY}{\tau_{eff}} = \frac{0.017}{5.2x10^{-9}s} = 3.3x10^6 s^{-1}$$

$$\gamma_{nonrad} = \frac{\gamma_{rad}(1-QY)}{QY} = \frac{3.3x10^6 s^{-1} \times (1-0.017)}{0.017} = 189x10^6 s^{-1}$$

*Determination of the number of clusters in optical probe volume* – The optically probed volume is cylindrical in shape with a height (cuvette thickness), $h$, of 1.0 cm and a diameter, $D_{opt} = 2r_{opt}$, of 0.5 cm. This corresponds to a volume of 0.79 mL.

$$V_{opt} = \pi r_{opt}^2 h = \pi \frac{(0.5\ cm)^2}{4} \times 1.0\ cm = 0.785\ cm^3 = 0.79\ mL \qquad [4a]$$

At a cluster concentration, $C_{cluster}$, of 31 ug of cluster / 1 mL of hexane (31 ug/mL), the optical density of the peak at 324 nm is 0.6. The number of clusters can be determined from three separate methods.

**Method 1:** An approximation for the number of clusters uses an effective bulk density, $\rho_{eff}$, of the cluster, the size of the cluster, and the cluster concentration. The cluster radius, $r$, is ~0.8 nm. From a previous study of the F313 and F324,(*4*) the inorganic mass fraction of the cluster is ~30% and the organic fraction (from oleate ligands) is ~70%.

$$\rho_{eff} = f_{inorg}\rho_{CdS} + f_{org}\rho_{Oleate} = 0.3 \times 4.82\ g \cdot mL^{-1} + 0.7 \times 0.90\ g \cdot mL^{-1} \qquad [4b]$$
$$= 2.07\ g \cdot mL^{-1}$$

$$N = \frac{C_{cluster}V_{opt}}{\rho_{eff}V_{cluster}} = \frac{31 \times 10^{-6} g \cdot mL^{-1} \times 0.785\ mL}{2.07\ g \cdot mL^{-1} \times \frac{4}{3}\pi(0.8 \times 10^{-7} cm)^3} = 5.7 \times 10^{15} \quad [4c]$$

**Method 2:** Using the formula weight of the cluster, $Cd_{37}S_{20}(oleate)_{37}$ ($Cd_{40}S_{20}(oleate)_{40}$, rounded), and the cluster concentration, we can estimate the number of clusters in the optical probe volume. The formula weight of the cluster, $F_{W,cluster}$, is 15,253 g mol$^{-1}$ (16,438 g mol$^{-1}$), which yields $9.6 \times 10^{14}$ ($8.9 \times 10^{14}$) clusters. We determined the formula weight of the cluster based on the inorganic composition ($Cd_{37}S_{20}$) that was used to fit the PDF. An earlier investigation on the F324 and F313 clusters showed the inorganic-organic composition as $[(CdS)(Cd(oleate)_2)]_x$, but the value of x was unknown.[7] Combining the PDF with the inorganic-organic composition, we expect the formula weight of the cluster to be $Cd_{37}S_{20}Oleate_{37}$.

$$N = \frac{C_{cluster}}{F_{W,cluster}} V_{opt} N_A = \frac{31 \times 10^{-6} g \cdot mL^{-1}}{15,253\ g \cdot mol^{-1}} \times 0.785\ mL \times 6.022 \times 10^{23} \frac{\#}{mol} \quad [4d]$$
$$= 9.6 \times 10^{14}$$

**Method 3:** Using the empirical fitting function from the work by W. Yu and coworkers(43), we can estimate the number of clusters from the peak position and optical density of the first absorption peak. The diameter of the cluster, $D$, is determined by the following polynomial(43),

$$D = (-6.65 \times 10^{-8})\lambda^3 + (1.96 \times 10^{-4})\lambda^2 - (9.24 \times 10^{-2})\lambda + (13.29) \quad [4e]$$
$$D = (-6.65 \times 10^{-8}) \times (324\ nm)^3 + (1.96 \times 10^{-4}) \times (324\ nm)^2 - (9.24 \times 10^{-2}) \times 324 nm + (13.29) = 1.67\ nm$$

where, $\lambda$ is the wavelength of the first absorption peak. The extinction coefficient, $\varepsilon$, is determined from the cluster diameter by the following polynomial(43),

$$\varepsilon = 21536 \times D^{2.3} = 21536 \times 1.67^{2.3} = 7.0 \times 10^4\ M^{-1} \cdot cm^{-1} \quad [4f]$$

where the units of M are moles of clusters per liter of solvent. The number of clusters can be estimated from the Beer-Lambert law,

$$A = \varepsilon CL \quad [4g]$$

where $C$ is the cluster concentration and $L$ is the path length of the optical beam through the sample. By rewriting $C$ in terms of cluster number,

$$C = \frac{N}{N_A V_{opt}} \quad [4h]$$

where $N_A$ is Avogadro's Number and solving for $N$ in the Beer-Lambert's law, the number of clusters is $4 \times 10^{15}$.

$$N = \frac{A}{\varepsilon L} V_{opt} N_A \quad [4i]$$

$$N = \frac{0.6}{7.0 \times 10^4 M^{-1} \cdot cm^{-1} \times 1\ cm} \times 0.785\ cm^3 \times 6.022 \times 10^{23} \frac{\#}{mol}$$
$$= 4.1 \times 10^{18} \# \cdot \frac{cm^3}{L}$$
$$N = 4.1 \times 10^{18} \# \cdot \frac{cm^3}{L} \cdot \frac{L}{1000\ cm^3} \cong 4 \times 10^{15} \#$$

The number of clusters estimated from each method are consistent with each other to within the same order of magnitude: $10^{15}$ clusters.

*Lifetime Calculation* – To estimate the average lifetime of a transforming cluster, we use 1) the rate constant from the transformation kinetics or 2) the change in optical density, which is proportional to the change in the number of clusters, within a given time step. For these examples, we use the 70°C conversion experiment with a rate constant $k_c = 8.0x10^{-3}$ s$^{-1}$. With an initial optical density of ~0.5 for the first absorption peak, there are ~$10^{15}$ clusters. At early times in the conversion process at 70°C, the change in optical density, $\Delta OD_t$, is ~0.01 per 5 s time step, $t_{step}$, for which the fraction of clusters, $f_{trans}$, that have transformed is 0.002. This change in optical density equates to ~$2x10^{13}$ clusters that have transformed.

$$\Delta N = \frac{\Delta OD_t}{OD} N = f_{trans} N = 0.002 \times 10^{15} = 2.0x10^{13} \# \quad [5a]$$

Therefore, the average lifetime of the transitioning clusters within this 5 s time step is,

$$\tau_{trans,1} = = \frac{t_{step}}{\Delta N} = \frac{5\ s}{2x10^{13}} = 2.5x10^{-13} s \quad [5b]$$

An alternative approach uses the rate constant of the transformation and the initial number of clusters to determine the average lifetime of the transitioning cluster.

$$\tau_{trans,2} = \frac{1}{Nk_c} = \frac{1}{10^{15} \times 8x10^{-3} s^{-1}} = 1.2x10^{-13} s \quad [5c]$$

These two approaches yield similar lifetimes for the transforming cluster, and the lifetimes are on the order of bond vibration frequencies. Correspondingly, these lifetimes are also typical for unimolecular reactions traversing a transition state. Therefore, we suspect the clusters do not pass through a metastable intermediate as they transform. Further evidence toward this is based on the temporal evolution of the absorption spectra. From the fitting of the two pure component spectra for the F324 and F313 and the evolution of the total spectral area of the transformation, we can identify if an intermediate has accumulated. There is a first order exponential response for the evolution in the fraction of the F324 and the F313. We find the sum of these two fractions equate to ~1.0 over the transformation processes. Likewise, the normalized total spectral area remains constant (~1.0) when summed over the wavelengths 220-350 nm throughout the transformation process (see **Fig. S8B**). These two results indicate a constant cluster number and that the only species are F324 and F313. Hence, the clusters do not transition through a metastable intermediate.

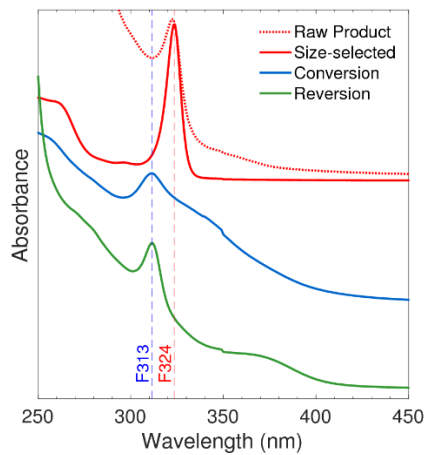

**Fig. S1A.**

Transformation cycle of clusters produced at dilute conditions. The raw product of the dilute synthesis is impure and contains large nanocrystals. The MSCs were size-selectively precipitated. The size-selective cluster appears to be a pristine F324 species as in the concentrated synthesis, and F324 synthesized under dilute conditions converts to F313 upon exposure to alcohol, but larger nanocrystals are produced in the process. Removing the alcohol and heating induces nucleation and growth of more nanocrystals.

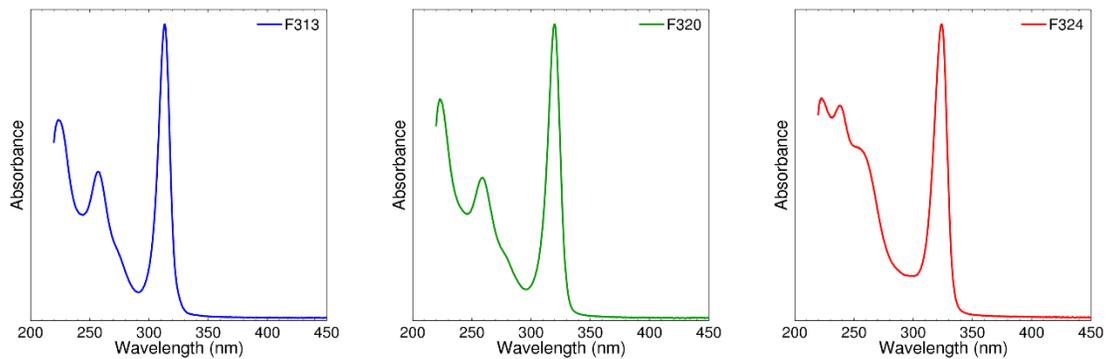

**Fig. S1B**

*Left* – Absorption spectrum of a pristine F313 thin film. *Middle* – Absorption spectrum of a pristine F320 thin film. *Right* – Absorption spectrum of a pristine F324 thin film.

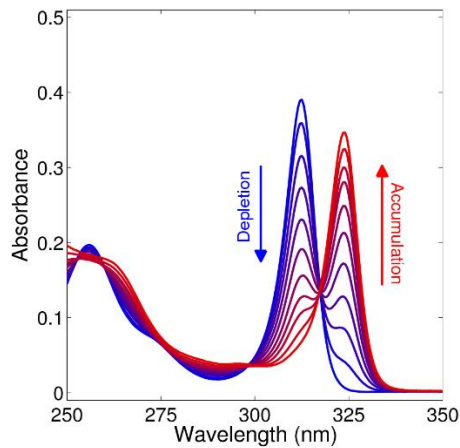

**Fig. S1C**
Ex-situ reversion process of clusters performed in a hexane. Here, F313 is dissolved in hexane with an initial optical density of 0.39 at 313 nm. The sample is placed into a hot oil bath at 70°C. Thereafter the sample is periodically taken out of the hot oil bath and rinsed, and its absorption spectrum is recorded at room temperature. The sample is heated and cooled repeatedly until complete reversion. The final optical density is 0.35 at 324 nm. The difference in optical density between a pure F313 and a pure F324 within a closed system would suggest either a change in cluster concentration or a change in extinction coefficient. Given that the difference in the optical densities of the excitonic peak is small (~0.04), a bimolecular collision of F313 clusters cannot produce the F324 cluster. Therefore, we attribute the difference in optical density to a change in extinction coefficient with a constant cluster concentration. The presence of multiple isosbestic (i.e., 299 nm and 318 nm) indicate a conversion between only 2 species.
<insert page break here>

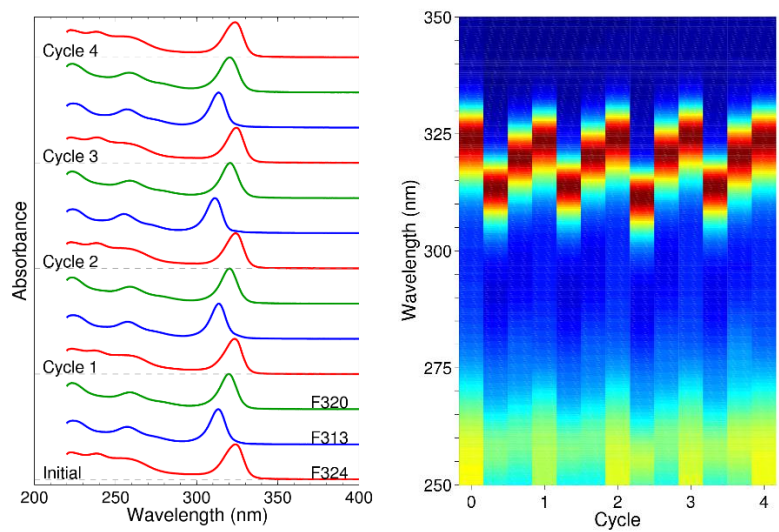

**Fig. S1D**

*Left* – Waterfall plot of the absorption spectra for the transformation cycle (**Fig. 2A**) with the respective F320 intermediate. *Right* – Contour map of the absorption spectra for the transformation cycles.

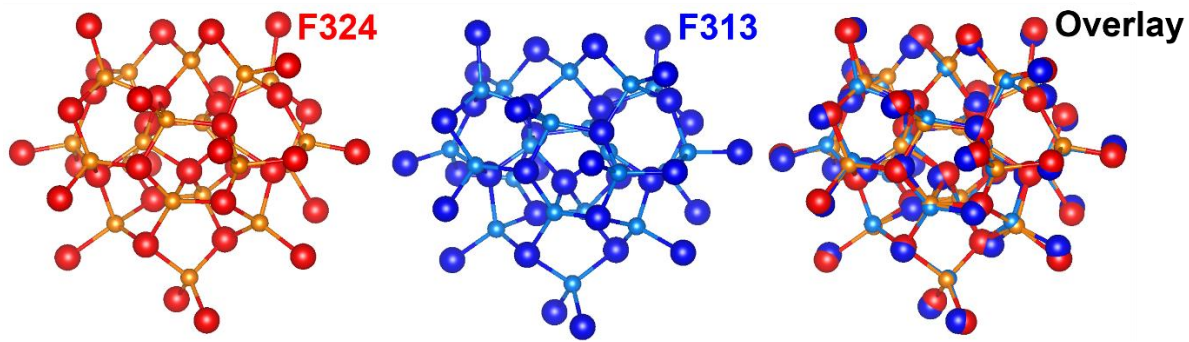

**Fig. S2A**

*Left* – Fitted F324 structured determined from the reverse Monte Carlo algorithm. The red spheres represent cadmium atoms and the orange spheres represent sulfur atoms. *Middle* – Fitted F313 structure determined from reverse Monte Carlo. The dark blue spheres represent cadmium atoms and the light blue spheres represent sulfur atoms. *Right* – The fitted F324 and F313 structures overlaid with respect to the same geometric center.

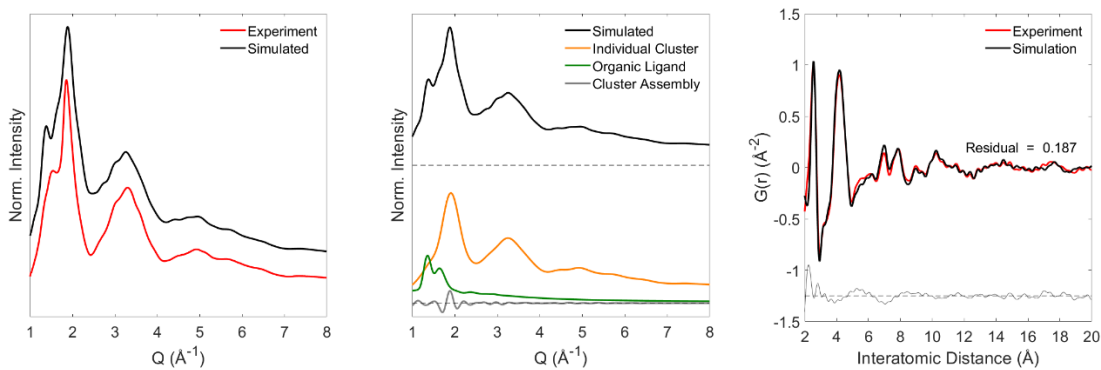

**Fig. S2B**

*Left* – Simulated XRD pattern of the F324 cluster (**Fig. S2A**) compared to the experimental XRD (total scattering) pattern from the synchrotron source. There is strong agreement between the simulation and the experimental data. There is a slight shift in the simulated peak at ~1.4 Å$^{-1}$ to lower Q, compared to the experimental peak. This peak is attributed the organic ligands and we calculate its scattering pattern directly from its single crystal data file (see **Methods** for details) without modification of its structure file. *Middle* – Scattering components that comprise of simulated XRD pattern. *Right* – Simulated PDF of the F324 cluster (**Fig. S2A**) compared to the experimental PDF. The simulated fit to the experimental data is a good fit (Residual <0.2). The broad oscillations at small interatomic distances relates to the small shift in the organic ligand peak from the XRD pattern. Top plot is PDF and bottom plot shows the difference between the PDFs plotted in the top plot.

S17

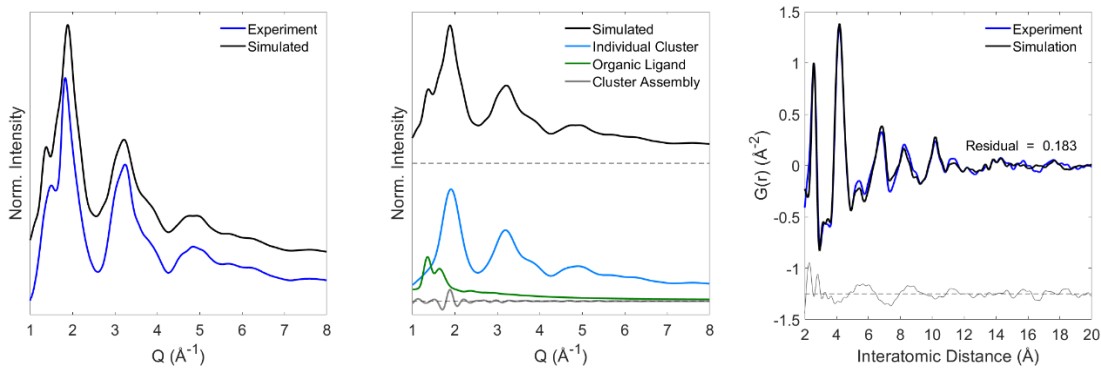

**Fig. S2C**

*Left* – Simulated XRD pattern of the F313 cluster (**Fig. S2A**) compared to the experimental XRD (total scattering) pattern from the synchrotron source. There is strong agreement between the simulation and the experimental data. There is a slight shift in the simulated peak at ~1.4 Å$^{-1}$ to lower Q, compared to the experimental peak. This peak is attributed the organic ligands and we calculate its scattering pattern directly from its single crystal data file (see **Methods** for details) without modification of its structure file. *Middle* – Scattering components that comprise of simulated XRD pattern. *Right* – Simulated PDF of the F313 cluster (**Fig. S2A**) compared to the experimental PDF. The simulated between the experimental is a good fit (Residual <0.2). The broad oscillations at small interatomic distances relates to the small shift in the organic ligand peak from the XRD pattern. Top plot is PDF and bottom plot shows the difference between the PDFs plotted in the top plot.

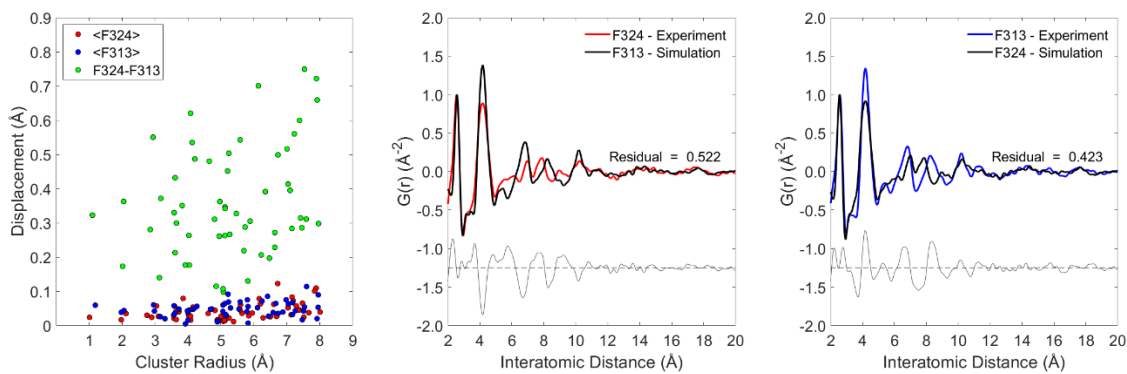

**Fig. S2D**

Statistics from repeated structure fits and reverse fits. *Left* – Atomic deviation (displacement) with respect to its radial position within the cluster for multiple structures that fit the two isomer PDFs. The <F324> and <F313> labels denote the mean deviation of individual atomic positions between different F324 and F313 structures, respectively, generated from repeated reverse MC algorithms. There are 20 F313 structures and 10 F324 structures generated from the reverse MC algorithm with residuals of <0.2 (when fitted to their respective PDFs). F324-F313 denotes the atomic displacement of the atoms between the best fit structure of the F324 and F313. The mean deviation between the individual atomic positions of the repeated structures (<F324> and <F313>) is ~0.04 Å and the atomic displacement between the best structures are, on average, ~0.40 Å. This order of magnitude increase in the atomic displacement between cluster structures (F324 and F313) compared to the repeated structures of the same cluster indicate that our F324 and F313 structures are unique and structures generated from the reverse MC algorithm are reproducible. *Middle* – PDF of the experimental F324 cluster compared to the PDF of the simulated F313 cluster. *Right* – PDF of the experimental F313 cluster compared to the PDF of the simulated F324 cluster. The residuals of both fits are very large (>0.2), which further supports our structures being unique and are not interchangeable. In PDF figures, top plot is PDF and bottom plot shows the difference between the PDFs plotted in the top plot.

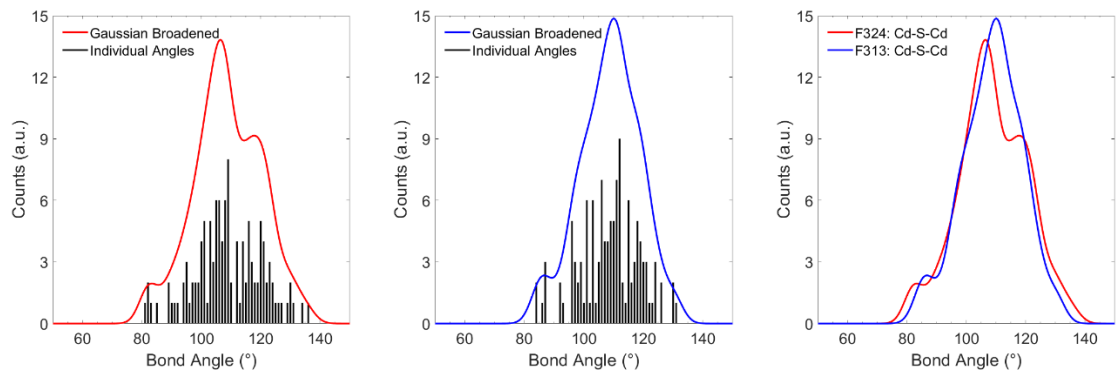

**Fig. S2E**

*Left* – Cd-S-Cd bond angles distribution (histogram) for the F324 cluster. There is a bimodal distribution (~105° and ~120°) of the Cd-S-Cd bond angles for the F324. *Middle* – Cd-S-Cd bond angles distribution (histogram) for the F313 cluster. There is a single distribution (~109.5°) of the Cd-S-Cd bond angles for the F313. *Right* – Overlay of the F324 and F313 Cd-S-Cd bond angle distribution. There is a slight broadening in the overall angle distribution in the F324 compared to the F313. The bond angles for both distributions are rounded to the nearest degree and then broadened with a normal distribution with a standard deviation of 3° for clarity.

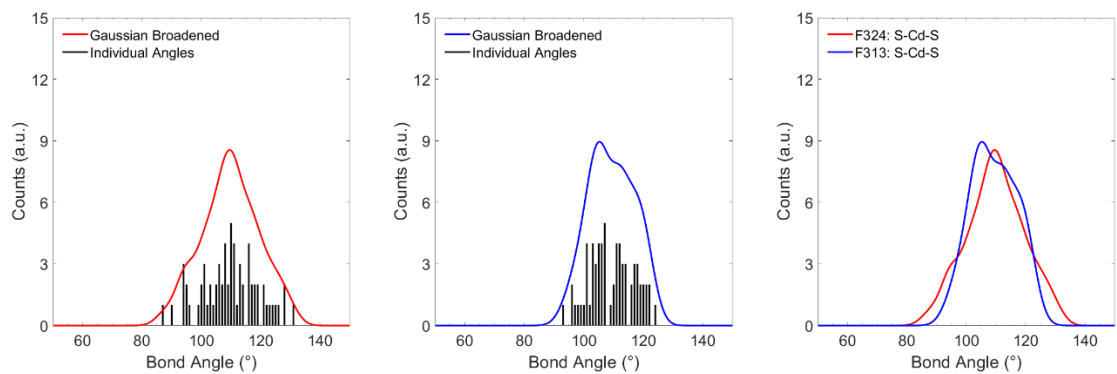

**Fig. S2F**

*Left* – S-Cd-S bond angles distribution (histogram) for the F324 cluster. There is a single distribution (~109.5°) of the S-Cd-S bond angles for the F324. *Middle* – S-Cd-S bond angles distribution (histogram) for the F313 cluster. There is a bimodal distribution (~105° and ~117°) of the S-Cd-S bond angles for the F313. *Right* – Overlay of the F324 and S-Cd-S bond angle distribution. There is a slight broadening in the overall angle distribution in the F324 compared to the F313. The bond angles for both distributions are rounded to the nearest degree and then broadened with a normal distribution with a standard deviation of 3° for clarity.

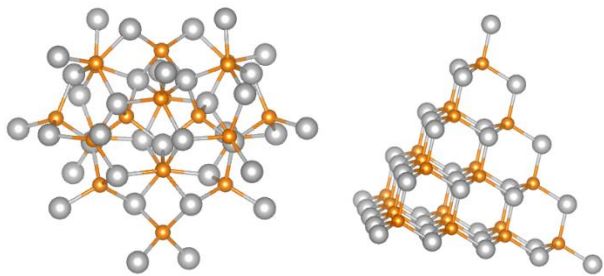

**Fig. S2G**

*Left* – An InP MSC structure from single crystal (*14*) data with the In and P atoms replaced with Cd and S atoms, respectively. *Right* – A CdS MSC tetrahedral structure with the composition of $Cd_{40}S_{20}$. This structure and variations of it have been previously reported for the CdS/Se clusters (*16*, *17*).

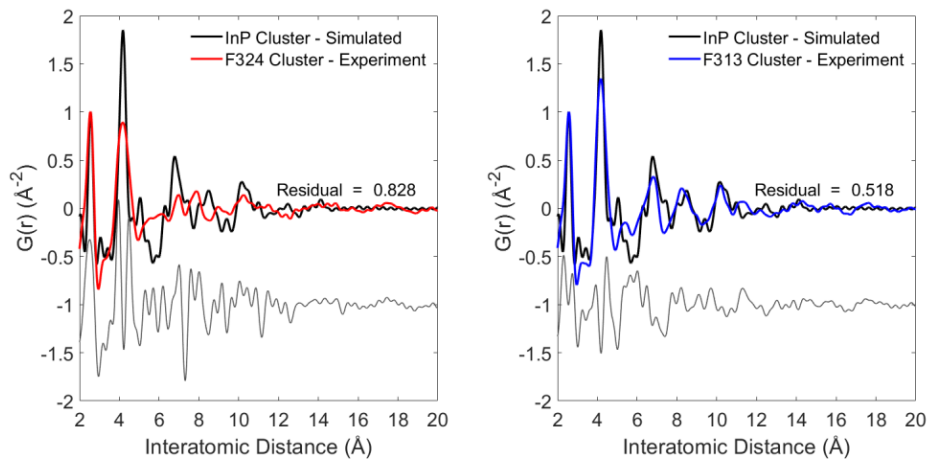

**Fig. S2H**

Comparison of the experimental cluster PDFs (*Left* – F324 and *Right* – F313) to the simulated PDFs for the original InP cluster (**Fig. S2G**) (*14*). There are large similarities between the peak positions of the F313 and the InP cluster. In PDF figures, top plot is PDF and bottom plot shows the difference between the PDFs plotted in the top plot.

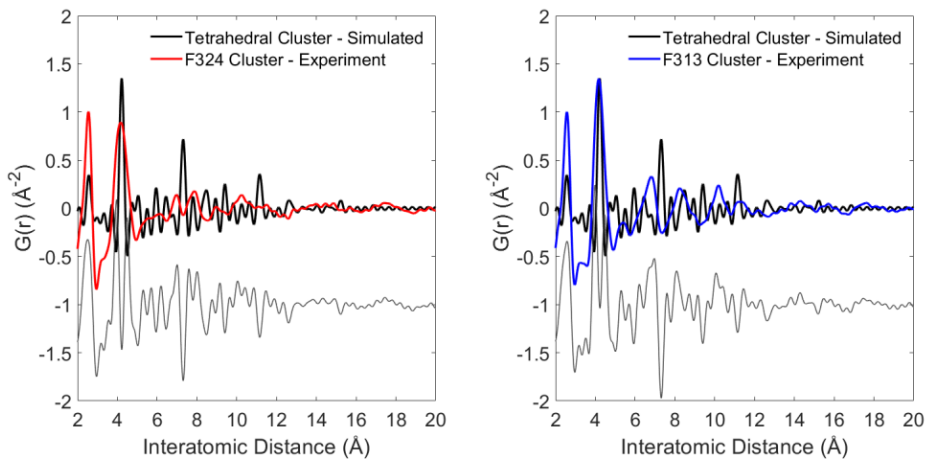

**Fig. S2I**

Comparison of the experimental cluster PDFs (*Left* – F324 and *Right* – F313) to the simulated PDFs for a CdS tetrahedral cluster (**Fig. S2G**). There are large differences in peak positions between the tetrahedral cluster and our experiment data from the F324 and F313. In specific, the ratio between the $1^{st}$ and $2^{nd}$ nearest are 1:3 in the tetrahedral, but nearly 1:1 in the F324 and F313, and there is a large definitive peak at ~7.5 Å in the tetrahedral cluster that is not present in the F324 and F313. When the reverse Monte Carlo algorithm is performed on the tetrahedral cluster, residuals do not fall below 0.2 and there are large structural deviations between repeat structures: this indicates that our clusters are not tetrahedron. Top plot is PDF and bottom plot shows the difference between the PDFs plotted in the top plot.

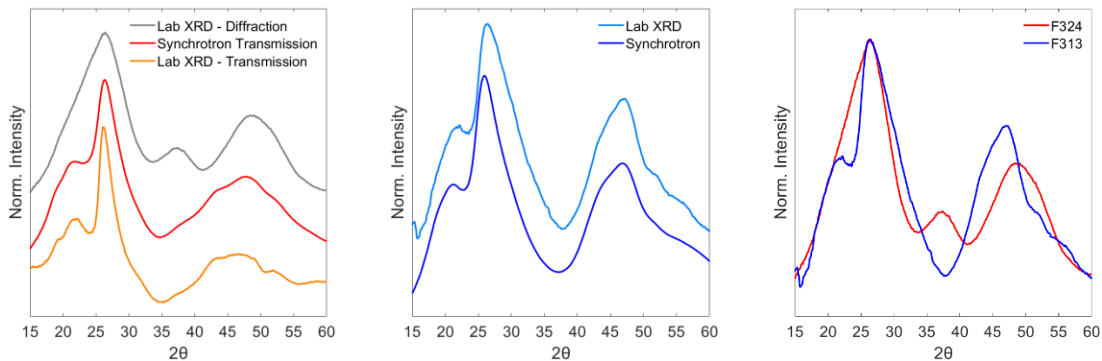

**Fig. S2J**

*Left* – XRD patterns of the F324 collected from a lab x-ray diffractometer and a synchrotron source (i.e., CHESS)(see **Methods**). Comparing the transmission mode of the lab XRD and the synchrotron XRD, the patterns mostly correspond. (The differences between synchrotron XRD patterns generated from a total scattering arrangement and patterns from lab-based XRD are primarily equipment related: coherency of the x-rays, beam size, data collection geometry, background intensity, bandwidth (monochromatic or not), divergence (or collimation) of the beam, etc.) However, there are more variations between the patterns from the two techniques (i.e., reflection vs. transmission) used to collect the patterns, which can be attributed to the sample perpetration methods that lead to different texturing. In specific, there is a pronounced peak at ~37° 2θ in diffraction/reflection mode, while this feature is a shoulder in transmission mode. Another difference is the peak at ~26° 2θ, which is narrower in transmission mode compared to diffraction mode. These differences are attributable to the mesophase that these clusters are formed within, as previously reported (*11*) which is causing texturing differences between the samples. The lab XRD patterns are collected from the same sample, but for the reflection geometry the sample is flat and level against a substrate and has a thickness of 2-3 mm, while for the transmission measurement the samples need to be thinner (200-500 um thickness) and sheared onto the holder substrate, which deforms the sample. *Middle* – XRD patterns of the F313 collected from a lab x-ray diffractometer and a synchrotron source (i.e., CHESS)(see **Methods**). There are large similarities in the diffraction patterns of the F313 in terms of relative peak intensities and position. *Right* – Overlay of F24 and F313 diffraction patterns collected from a lab x-ray diffractometer.

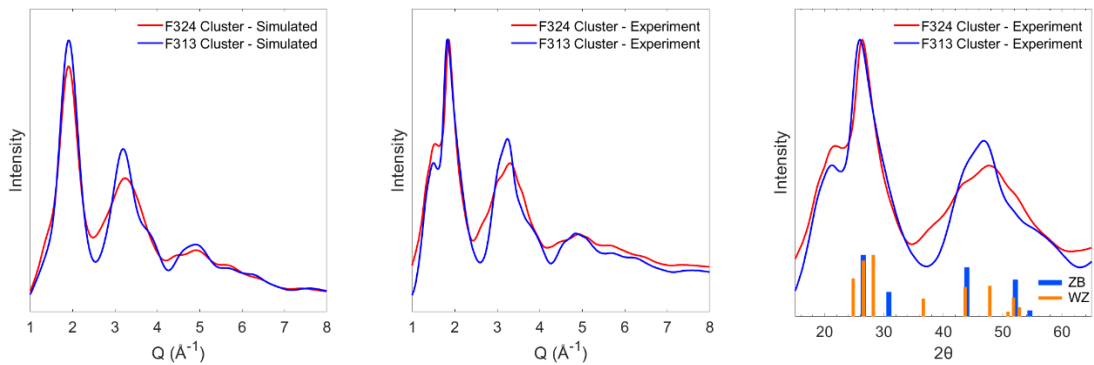

**Fig. S2K**

*Left* – Simulated XRD pattern of only the F324 and F313 clusters from the structure in **Fig. S2A**. The organic and mesophase component has been excluded from the simulation. *Middle* – Experimental XRD (total scattering) of the F324 and F313 clusters. *Right* – Experimental XRD (total scattering) of the F324 and F313 clusters with the scattering vector, Q, converted to 2θ using a λ of 1.54 Å (Cu K-α source). Theoretical peaks for zinc blende (ZB, PDF#00−010−0454) and wurtzite (WZ, PDF#00-041-1049) have been overlaid. The F324 resembles more of a WZ phase and the F313 resembles more of a ZB phase. The most distinguishing feature is the shoulder at ~36.6° 2θ in the F324 that matches a signature peak in the WZ phase.

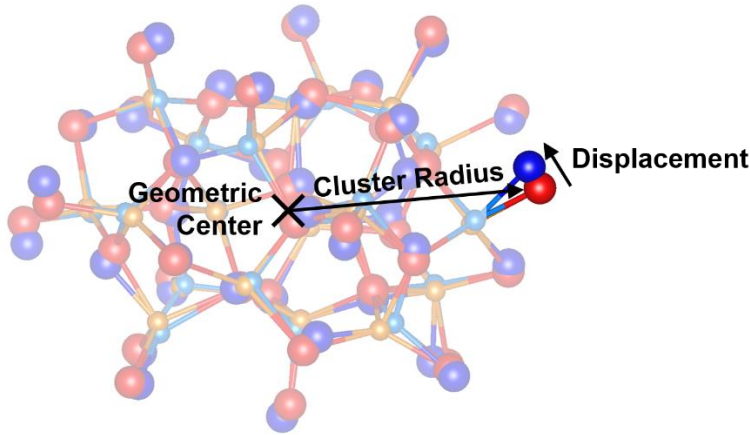

**Fig. S2L**

Illustration for the determination of atomic displacement between the F324 and F313 structure with respect to radial position. The cluster radius and displacement are Cartesian vector magnitude calculations. The vector magnitude for the radial positions is determined from the geometric center of the F324 to the position of each atom in the F324. The vector magnitude for the displacement is determined from the position of each atom in the F324 to its equivalent atom in the F313. As an example, we use cadmium atom #54 (not shaded atom) in the F324 and F313 structures in the following calculations.

**Radial Position of Cadmium Atom #54 in F324**

F324 Cartesian Coordinate: [1.15, -5.87, 5.14] Å
Geometric Center: [0.00, 0.00, 0.00] Å

$$Radial\ Position = \sqrt{(1.15 - 0.00)^2 + (-5.83 - 0.00)^2 + (5.14 - 0.00)^2} = 7.86\ \text{Å}$$

**Displacement of Cadmium Atom #54**

F324 Cartesian Coordinate: [1.15, -5.87, 5.14] Å
F313 Cartesian Coordinate: [0.65, -5.45, 5.42] Å

$$Displacement = \sqrt{(1.15 - (0.65))^2 + (-5.87 - (-5.45))^2 + (5.14 - (5.42))^2} = 0.71\ \text{Å}$$

$$Strain = \frac{Displacement}{Radial\ Position} = \frac{0.71\ \text{Å}}{7.86\ \text{Å}} \times 100\% = 9.0\%$$

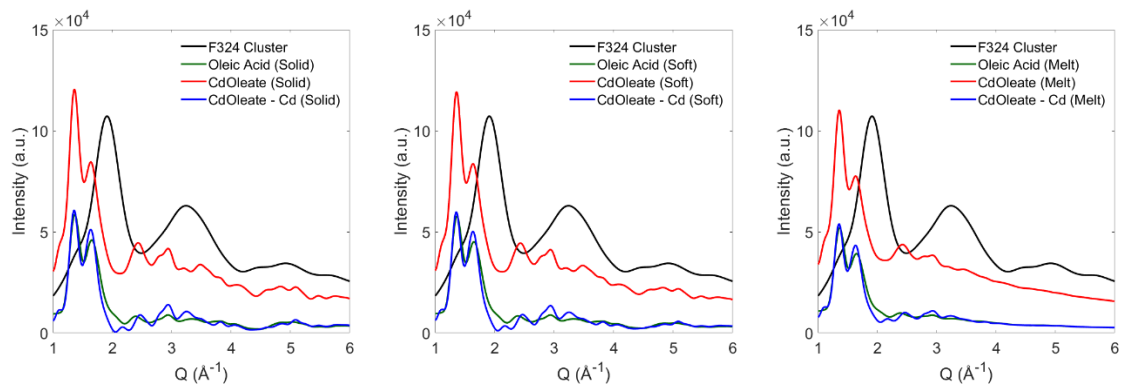

**Fig. S2M**

Simulated XRD patterns of the *Left* – solid, *Middle* – soft , and *Right* – melted (melt) organic ligands. The solid ligands are simulated from a single crystal structure with no thermal broadening (i.e., T = 0 K). The soft and melted structures have thermal broadening equivalent to a change in atomic position that is 10% and 30%, respectively, of the CdS bond length. As the thermal broadening increases, the intensity and features at high Q fade. Nearly 90% of the intensity in the CdOleate at high Q-space (> 2 Å$^{-1}$) derives from only the Cd-Cd interactions, which are non-existent in the CdS MSC inorganic core. All atom-atom interaction that involves organics atoms (i.e., C and O) have nearly negligible contribution to the scattering intensity at high Q-space.

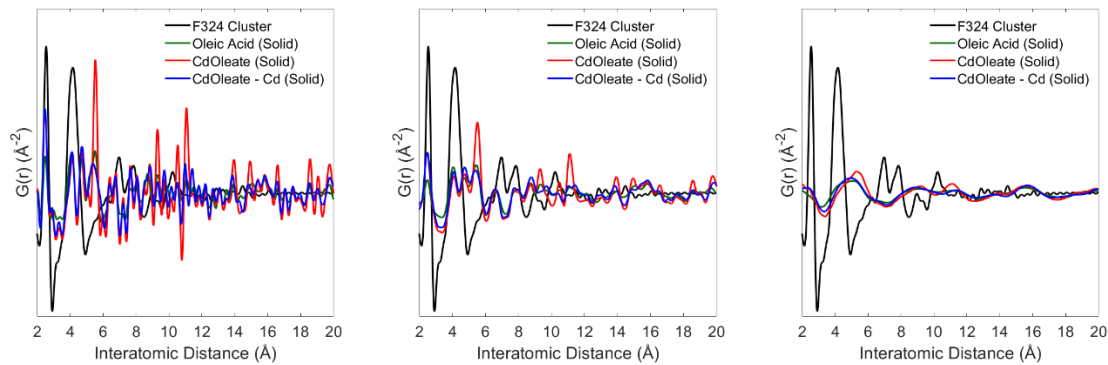

**Fig. S2N**

PDF of the *Left* – solid, *Middle* – soft, and *Right* – melted (melt) organics ligands from the simulated XRD patterns in **Fig. S2M**. The solid ligands have sharp features that propagate across all interatomic distances. These features create artificial peak in the PDF that do not resemble our experimental data. In the soft ligands, these features from the organics have broadened significantly and affect the amplitude and position of the cluster peaks in the PDF. The peak abundance of the simulated organic ligands is much greater than the experimental data, which indicates the ligands are likely to be amorphous. The melted ligands features have broadened substantially and the ligand contribution to the PDF, now only affects the peak amplitude (i.e., density of the material) and agrees with the data.

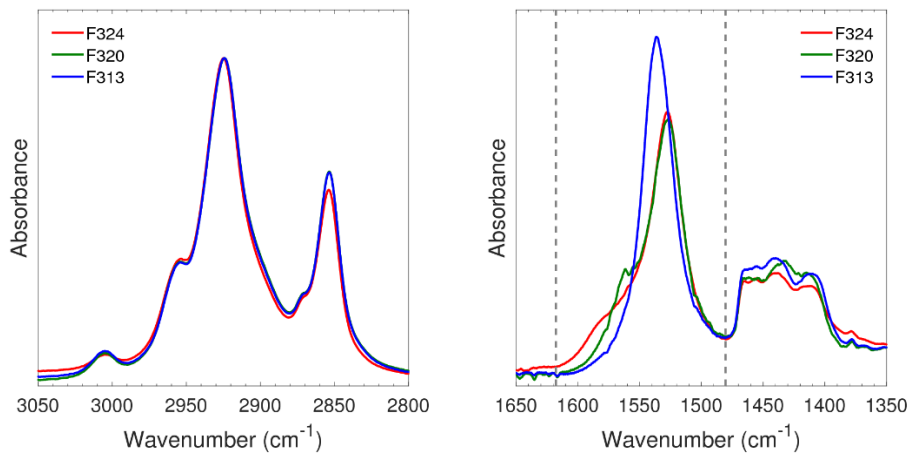

**Fig. S3A**

*Left* – FTIR spectra of the C-H stretching bands for the F313, F320, and F324 clusters. The spectra are normalized to the peak at 2925 cm$^{-1}$ (CH$_2$ asymmetric stretch). *Right* – FTIR spectra of the carboxylate stretching bands for the F313, F320, and F324 clusters. The dashed lines identify the region of carboxylate asymmetric stretches. The differences between spectral areas in this region between F313, F320, and F324 are small (2.5%), which indicates no changes in the number of bonds.

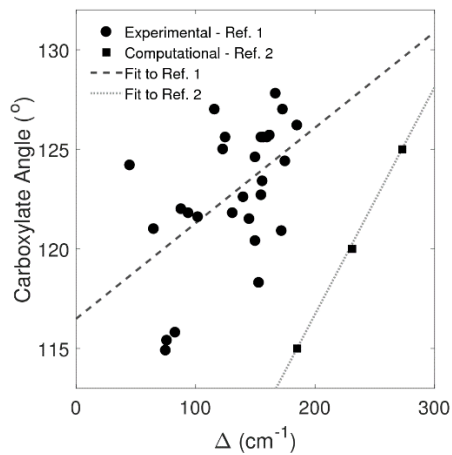

**Fig. S3B**

Relationship between the carboxylate bidentate angle ($\angle OCO$) to the difference ($\Delta$) between the symmetric and asymmetric stretch frequencies of the carboxylate. The values reported for Ref. 1 were measured experimentally from different transition metal acetates. The values reported for Ref. 2 are calculation from frequency shifts of a free acetate ion.
Fit for Ref. 1 (*24*): $\angle OCO = 0.048\Delta + 116.5$
Fit for Ref. 2 (*23*): $\angle OCO = 0.114\Delta + 93.9$

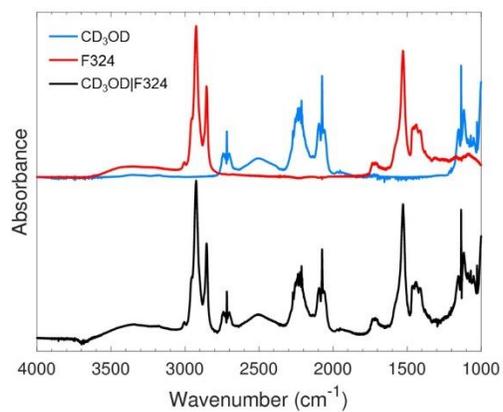

**Fig. S3C**

*Top* – FTIR spectra of deuterated methanol ($CD_3OD$) and F324 measured independently in a vapor cell. *Bottom* – FTIR spectra of F324 measured simultaneously with $CD_3OD$ vapor. The $CD_3OD$ does not share any vibrational features with the F324, which enables characterization of peak shifts.

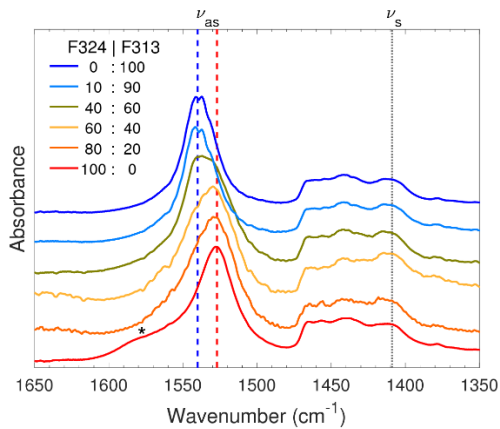

**Fig. S3D**

*Ex-situ* FTIR spectra of conversion process using deuterated methanol (CD$_3$OD) to initiate the transformation. UV-Vis absorption can be used to identify the ratio of F324 to F313. When the CD$_3$OD vapor is added to the cell, the shoulder (*) at 1580 cm$^{-1}$ disappears. As the F324 is converted to the F313, the 1528 cm$^{-1}$ peak diminishes and the 1540 cm$^{-1}$ peak increases proportional to the conversion fraction. There is no shift in the peak at 1409 cm$^{-1}$. The features at 1580, 1540, and 1528 cm$^{-1}$ are related to the asymmetric stretches of the carboxylate ($\nu_{as}$) and the peak at 1409 cm$^{-1}$ is the symmetric stretch of the carboxylate ($\nu_s$).

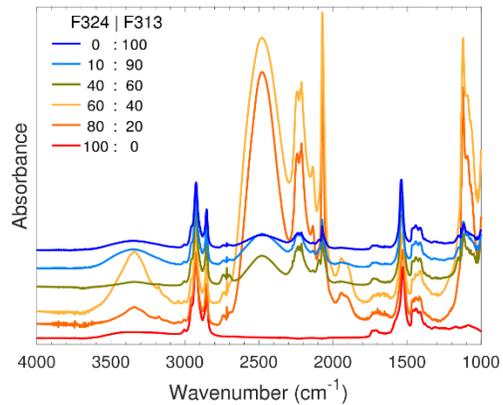

**Fig. S3E**

Full spectra of the ex-*situ* conversion process using deuterated methanol ($CD_3OD$). The $CD_3OD$ concentration changes periodically as a result of evaporation and condensation within the vapor cell, preventing quantitative analysis of how the deuterated methanol interacts with the film. However, the large changes in the $CD_3OD$ spectra do not influence the cluster spectra.

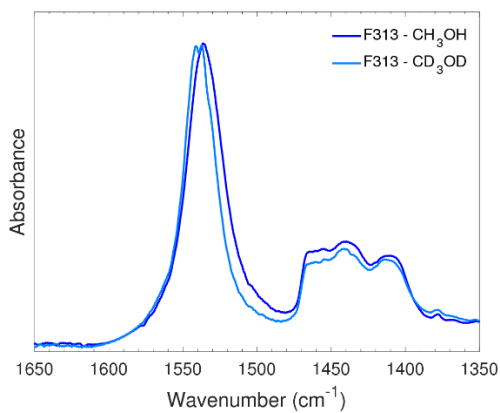

**Fig. S3F**

Comparison of F313 made with methanol with natural D abundance ($CH_3OH$) to F313 made with deuterated methanol ($CD_3OD$). The features between 1350 and 1480 cm$^{-1}$ are similar. The asymmetric carboxylate stretches for the $CH_3OH$ and $CD_3OD$ are at 1538 and 1540 cm$^{-1}$, respectively. This slight shift between the two could be a result of the decreased full-width half maximum of the F313 made with $CD_3OD$ and/or temperature fluctuations as observed with the exciton (see **Fig. S6**). We do not believe the shift is related to the difference in the coupling of the carboxylate of the ligand to the hydroxyl between the alcohols. With respect to $CH_3OH$, $CD_3OD$ interactions with the carboxylate should be shifted to lower frequencies, but neither of these effects are observed with the F313 made with $CD_3OD$.

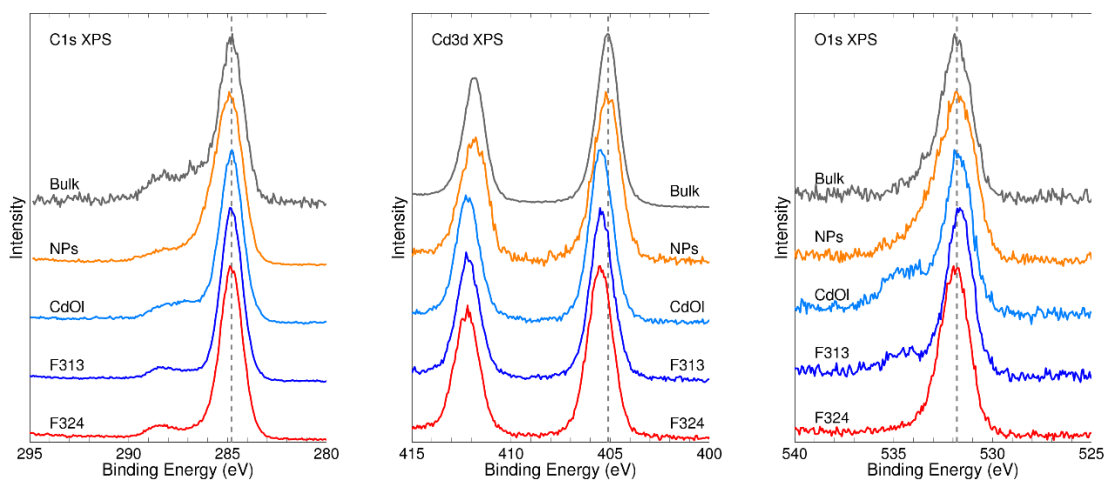

**Fig. S4A**

Stacked XPS spectra of F324, F313, cadmium oleate (Cd(oleate)), large cadmium sulfide nanoparticles (NPs), and bulk cadmium sulfide (Bulk). *Left* – Carbon 1s XPS spectra. The dotted line at 284.8 eV represents adventitious carbon. This peak is used as a reference for each sample. *Middle* – Cadmium 3d XPS spectra. The dotted line at 405.1 eV represents the Cd $3d_{5/2}$ level for the $Cd^{2+}$ oxidation state. The bulk and NP spectra match that of the reference (405.1 eV). The F324, F313, and Cd(oleate) peaks are shifted to 405.5 eV. *Right* – Oxygen 1s XPS spectra. The dotted line at 531.8 eV is used to illustrate that the F324, Cd(oleate), NPs, and bulk samples have the same peak position. The F313 sample has the main peak shifted to 531.6 eV. The F313 and Cd(oleate) have a second peak at 534.5 eV, which indicate different oxygen species or binding configurations.

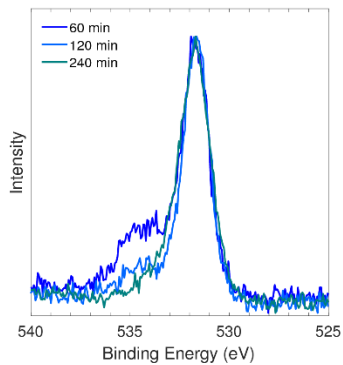

**Fig. S4B**

Evolution of the O1s XPS spectra for the F313. While the F313 is under vacuum, the peak at 535 eV diminishes with time, indicating the amount of methanol is decreasing. The XPS spectra are collected under ultra-high vacuum, which may cause desorption of weakly adsorbed species, such as the physisorbed methanol. It takes roughly 60 min to evacuate the XPS instrument. All O1s spectra are referenced to the 284.8 eV peak of the C1s spectra.

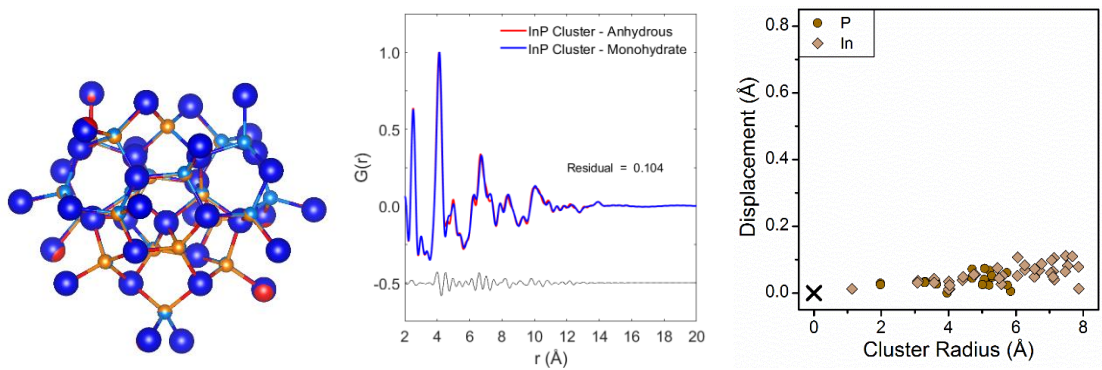

**Fig. S5**

Analysis of single crystal XRDs of InP clusters reported by Gary et al. (*14*). The crystallographic data of the InP clusters used in this analysis are reported in the Cambridge Crystallographic Database: deposition numbers are 1417965 (anhydrous) and 1417966 (monohydrate). *Left* – Overlay of the anhydrous (red) and a monohydrate (blue) InP cluster. Dark blue and red represent In atoms and the light blue and orange represent P atoms. Although the single crystal unit cell parameters differ, both unit cells are triclinic and the overlay of the InP cluster structures are nearly indistinguishable. *Middle* – PDF of the anhydrous and monohydrate cluster. The difference (black) between the two clusters is noise with a residual of 0.104. In the PDF figure, top plot is PDF and bottom plot shows the difference between the PDFs plotted in the top plot. *Right* – Atomic displacement of the atoms between the anhydrous and monohydrate structures with the cluster radius being that of the anhydrous structure. The displacement between the atoms are small and the subtle increase in displacement as the cluster radius increases may be due to the misorientation between the two clusters; within the single crystal the two clusters are oriented differently from one another and require a 3-D rotation around the geometric center to align the together. The presence of water in these structures has no influence on the inorganic structure.

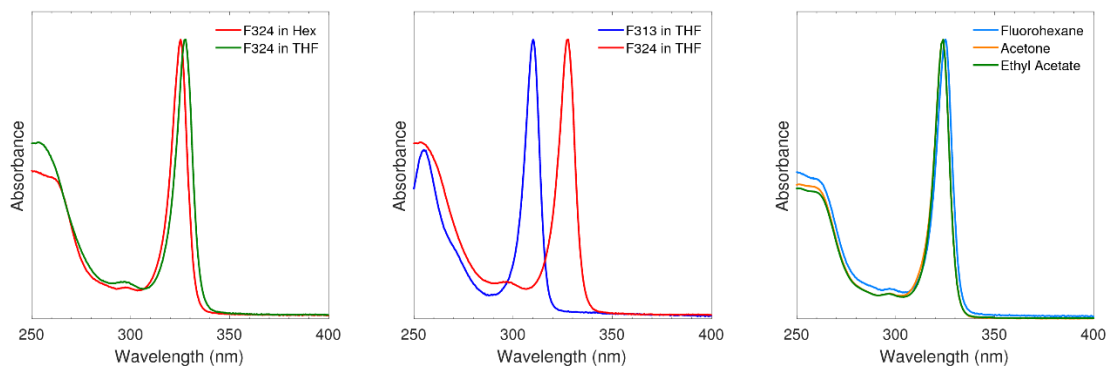

**Fig. S6A**

*Dielectric effects – Left –* Absorption spectra of F324 dissolved in hexane (Hex) and tetrahydrofuran (THF). Of solvents tested here that can dissolve the MSCs, THF has the largest dielectric constant. At ambient temperature, F324 in hexane has an excitonic peak at 324.0 nm, whereas in THF it has an excitonic peak at 327.5 nm. *Middle –* Absorption spectra of F324 and F313 dissolved in THF. The F313 isomer dissolved in THF has an excitonic peak at 310 nm. The separation energy between F324 and F313 in THF is 213 meV, whereas this energy separation is 135 meV in hexane. An increase in the dielectric constant increases the energy separation between the excitonic peaks for two pristine MSCs. *Right –* Absorption spectra of F324 cleaned with varying dielectric solvents and dissolved in hexane (small peaks shifts are due to temperature variations, see **Fig. S6D**). Cleaning the F324 with an aprotic solvent having low (fluorohexane) or high (ethyl acetate, acetone) dielectric constant does not substantially affect excitonic peak position. Only protic alcohols, whether high (methanol) or low (octanol) dielectric constant, can induce a F324→F313 transformation. Thus, MSC transformation is not the result of a change in dielectric environment.

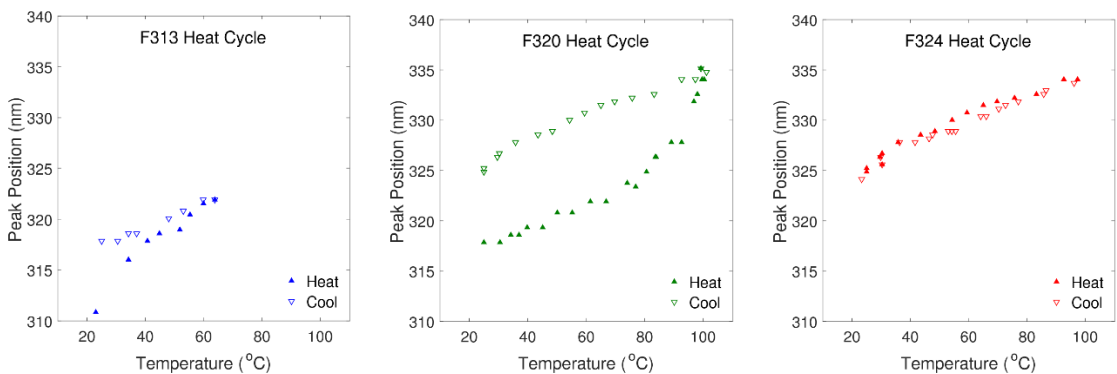

**Fig. S6B**

Thermal cycling of cluster films. Each cluster film has a strongly temperature dependent excitonic peak position. An F313 film will quickly transform to F320 when heated in dry atmosphere. A F320 film begins to revert to F324 when heated above 65°C. The F324 excitonic peak position is fully reversible with changes in temperature.

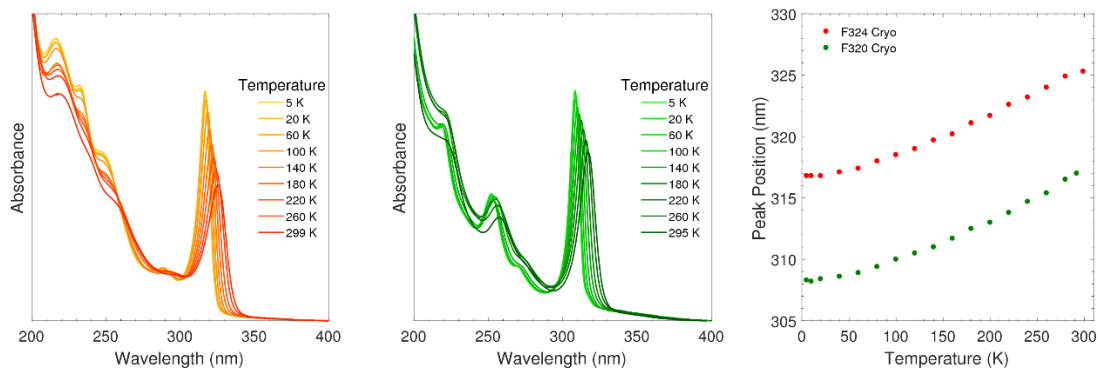

**Fig. S6C**

*Left* – Absorption spectra of MSC films below ambient temperature in a cryostat. *Middle* – Absorption spectra of F320 (originally F313, but dried under vacuum; see **Methods**) films. Both F324 and F320 films are slightly translucent, the scattering from which contributes to the sloping baseline observed. *Right* – Temperature-dependent excitonic peak positions of F324 and F320. The ~106 meV separation between them is constant.

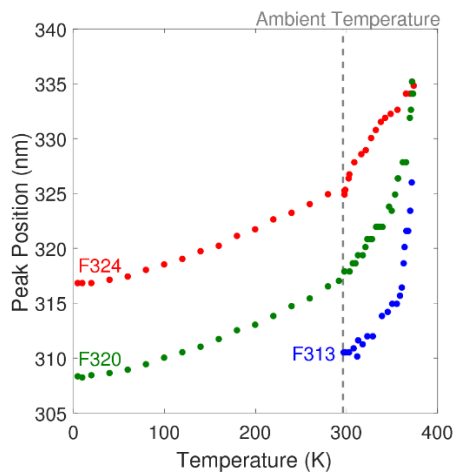

**Fig. S6D**

Influence of temperature on the excitonic peak position of MSC films. The F320 and F324 films are under dry environment (N$_2$ or He gas). The F313 film is under saturated methanol vapor. Removal of the methanol vapor causes the F313 to shift to F320 (see **Fig. S6B**). Peak positions below ambient were measured in a cryostat using the Cary 5000 spectrometer and peak positions at and above ambient were measured using the Ocean Optics UV-Vis spectrometer (see **Methods**). The F320 begins to revert to F324 at 65°C, and the F313 is unstable at 90°C (above the boiling point of methanol), reverting to F324.

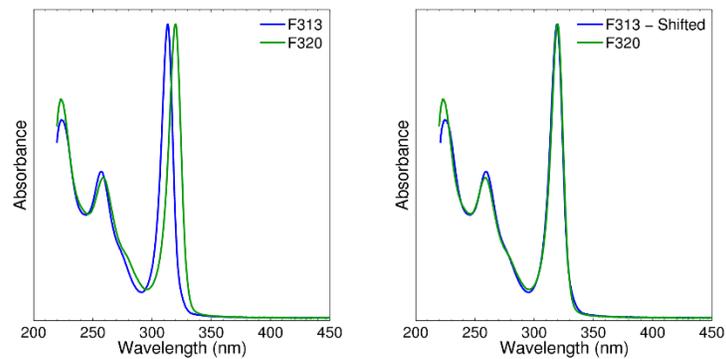

**Fig. S7A**

*Left* – Overlaid absorption spectra of F313 and F320. *Right* – Overlaid absorption spectra of F313 and F320, in which the wavelength of the F313 has a polynomial shift of $2x10^{-12} \times \lambda^5 + \lambda$, where λ is the original wavelength value.

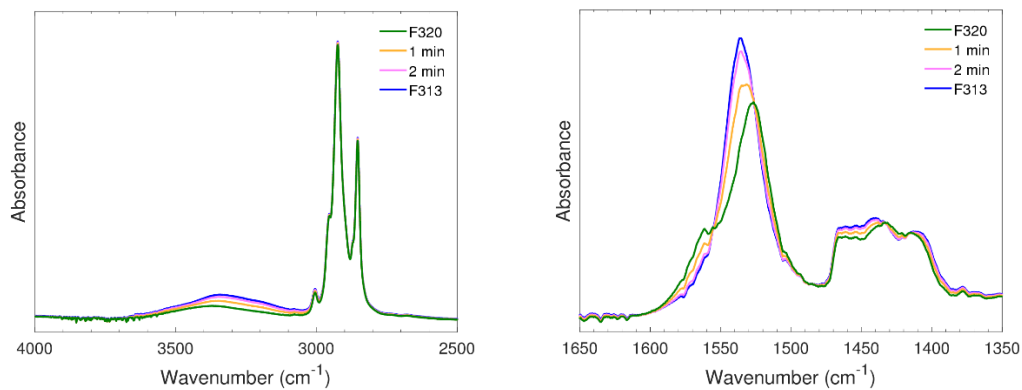

**Fig. S7B**

*Left* – FTIR spectra of the O-H (3100-3700 cm$^{-1}$) and C-H (2700-3100 cm$^{-1}$) stretches for the transition from F320 to F313. While under N$_2$, the F320 is stable. Upon exposure to ambient air, the F320 transforms back into the F313. The C-H stretches do not alter during the shift. There is an increase in intensity of the O-H stretch as the F320 shifts back to the F313, indicating the absorption of moisture from the air. *Right* – FTIR spectra of the carboxylate stretches for the transition from F320 to F313. The F320 has a strong peak at 1528 cm$^{-1}$ and a shoulder at 1560 cm$^{-1}$. Upon exposure to air, the strong peak of the F320 shifts to 1538 cm$^{-1}$ and the shoulder disappears. See **Fig. S7C** for the spectral area analysis.

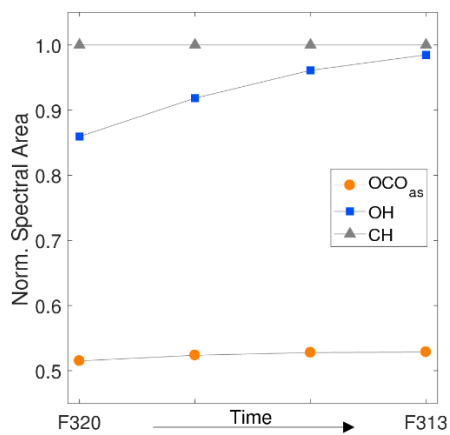

**Fig. S7C**

Spectral areas of the carboxylate asymmetric, O-H, and C-H stretches. The spectral areas are normalized with respect to the C-H stretches. The carboxylate spectral area does not change as the F320 shifts back to the F313, suggesting all the bidentate bonds are preserved and only the bond angles change. There is an increase in the O-H spectra area, which indicates there is more moisture in the film and that the F320 may be hygroscopic. We do not believe the increase in O-H stretch intensity is due to the $N_2$ being replaced with ambient air inside the vapor cell, since there is no significant difference between the vapor cell backgrounds whether the cell is filled with ambient air or $N_2$.

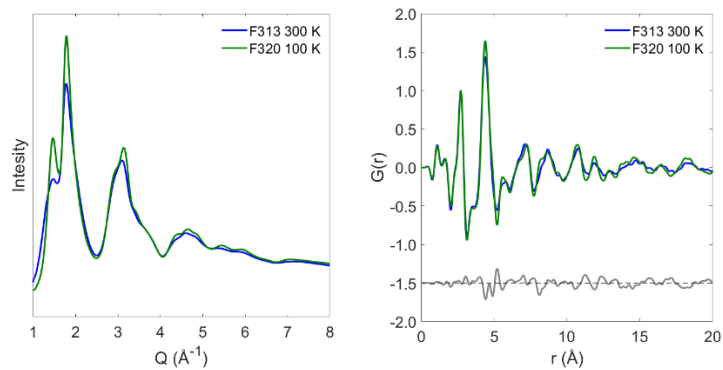

**Fig. S7D**

*Left* – Total scattering (XRD) pattern of the F313 at 300 K and the F320 at 100 K. The patterns are the same sample at different temperatures. Cryogenic $N_2$ vapor is passed over the F313 sample to cool the sample from 300 K to 100 K, which purges the air and dehumidifies the local area around the sample. This dehumidification mimics the conditions that produce the F320 as observed in the UV-Vis and FTIR spectra. Therefore, under the cryogenic conditions (temperature 100 K, $N_2$ vapor) the F320 is stabilized, and under ambient conditions and environment, the F313 is stabilized. The two scattering patterns correspond very well, with the exception peak narrowing, which is to be expected since thermal broadening is being reduced. *Right* – PDF of the F313 at 300 K and F320 at 100 K. The two PDFs are nearly identical and indicate that the F313 and F320 have the same structure.

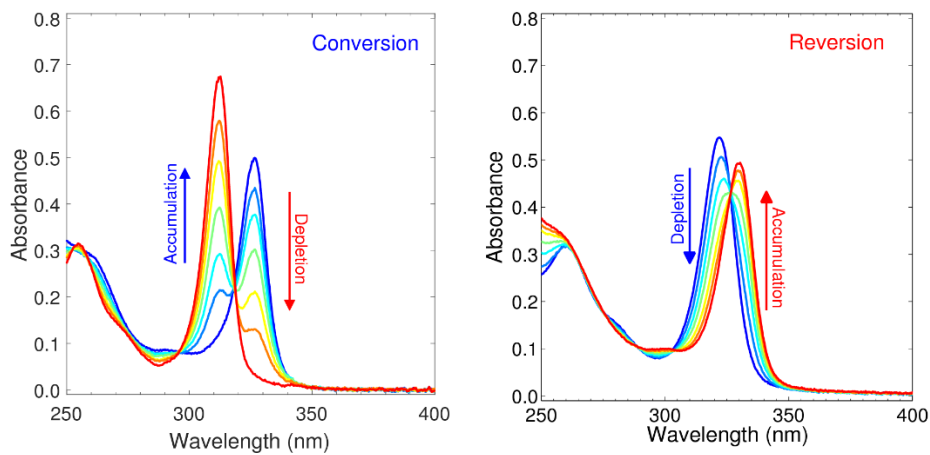

**Fig. S8A**

*Left* – Typical evolution of in-situ absorption spectra for the conversion process of a pristine F324 cluster thin film into F313 upon addition of alcohol. Temperature is 60°C. *Right* – Typical evolution of in-situ absorption spectra for the reversion of a pristine F313 cluster thin film to F320 upon removal of alcohol, followed by transforming to F324 upon heating. Temperature is 80°C.

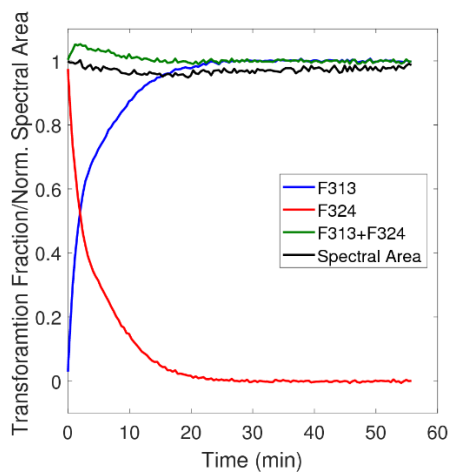

**Fig. S8B**

Conversion process for an MSC film at 60°C showing the evolution of the F324 and F313 fractional components and their sum. The spectral area, which is nearly constant throughout the transformation, is the sum of the absorbance between the wavelengths 257-350 nm and is normalized to the initial spectral area at t = 0 s. The sum of F324 and F313 fractions and the normalized spectral area are nearly 1.0 throughout the transformation, which indicates that there are no intermediates (F324 and F313 are the only species), and that F313 is produced one-for-one from every F324 consumed.

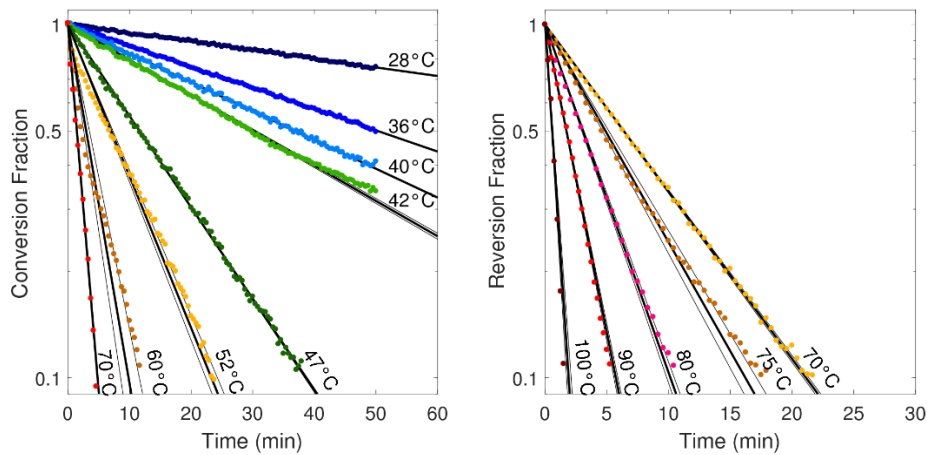

**Fig. S8C**

*Left* – Rate of conversion at various temperatures with kinetic fits to 90% conversion. *Right* – Rate of reversion at various temperatures with kinetic fits to 90% of reversion. The thick lines are fits to the data with the function form of $f_{324}(t)=e^{-kt}$, where $f_{324}(t)$ is the converted and reverted fraction at time *t*, and *k* is the rate constant and only fitting parameter. The thin lines are given by the estimated error in the rate constants. Error analysis of the least squares fitting is also performed (see **Tables S4**). Representative absorption spectra are shown in **Fig. S8A**.

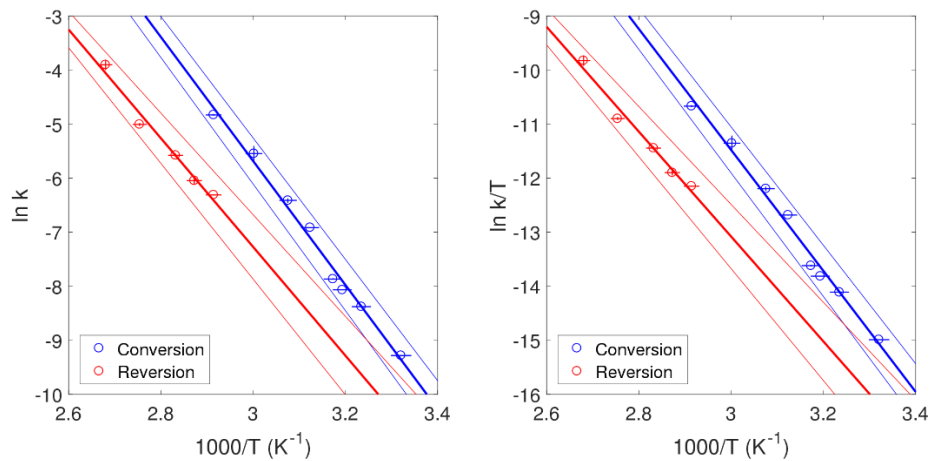

**Fig. S8D**

*Left* – Arrhenius plot of the rate constants from **Fig. S8C**. *Right* – Eyring plot of the rate constants from **Fig. S8C** normalized by temperature. Vertical error bars are error in the rate constant and may be smaller than the size of the markers (see **Table S4** for values). The estimated accuracy of the temperature (horizontal error bars) is ±2.0°C. Temperature fluctuations during measurement were ~±0.5°C. Temperature changes of the film throughout the transformation process give rise to shifts in the excitonic peak (**Fig. S6**) which are still smaller than the wavelength resolution of the instrument. Thick lines are the fit using the Arrhenius (left) or the Eyring (right) function, and the thin lines give the bounds determined by the uncertainty in the fitting parameters (see **Table S5,6** for values).

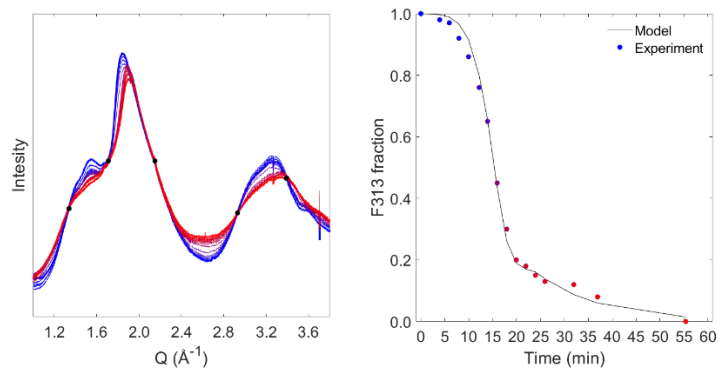

**Fig. S8E**

*Left* – In situ experimental XRD patterns of the transformation from F313 to F324 using the Cornell High Energy Synchrotron source. *Right* – The evolution of the F313 cluster in the reversion process. The fractions of the transformation are determined from a convex combination of the pure F313 and F324 XRD patterns (see **Fig. S8F** for fits and **Fig. S8G** for determination of fractional conversion). The experimental data is fit with 1st order reaction kinetics that has a time-temperature dependence.

$$\text{Fitting Function:} \qquad f_{313}(t) = 1 - e^{-kt}$$

This transformation is not isothermal (temperature varies with time, T(t)), and therefore the rate constant ($k(t,T)$) varies with transformation time. The rate constants are determined using the Arrhenius equation:

$$\text{Arrhenius Equation:} \qquad ln|k(t,T)| = lnA - \frac{E_a}{k_b T(t)}$$

with an activation energy of 0.87 eV and a prefactor of $8 \times 10^8$ s$^{-1}$. These kinetic parameters are in remarkable agreement with those determined from the spectroscopic kinetic study, which has an activation energy of 0.87 eV and a prefactor of $1.2 \times 10^9$ s$^{-1}$. The difference between the prefactors are within the experimental error of the spectroscopic kinetic study (see **Table S5**)

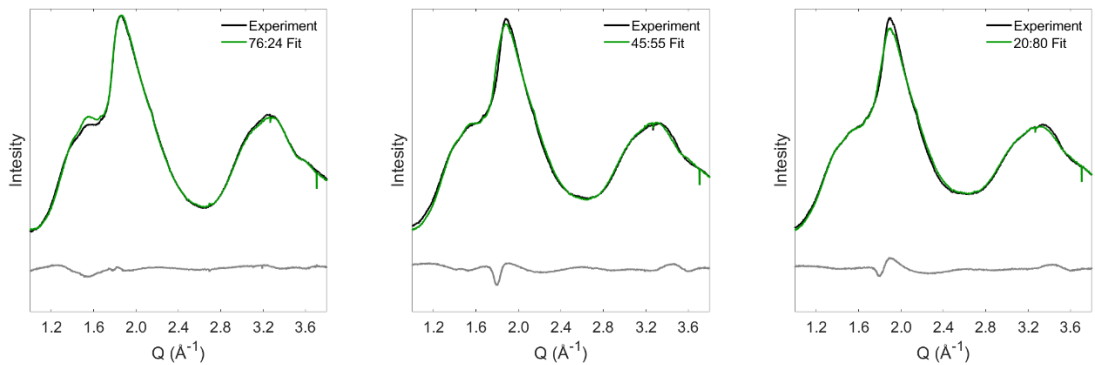

**Fig. S8F**

Representative fits to the in situ x-ray scattering F313 to F324 transformation. *Left* – 76:24 (F313:F324), *Middle* – 45:55 (F313:F324), and *Right* – 20:80 (F313:F324), ratio of the F313 and F324 pure components, respectively. The residual of all the fraction fits using the pure component F313 and F324 XRD patterns are good and <0.1. The majority of the residual derives from the small distortion around 1.5-2.0 Å$^{-1}$, which this distortion shifts to higher 2θ at higher temperatures (independent of transformation fraction). This feature likely relates to the mesophase assembly (texturing), as organics in this mesophase are much more sensitive to changes in x-ray scattering with temperature.

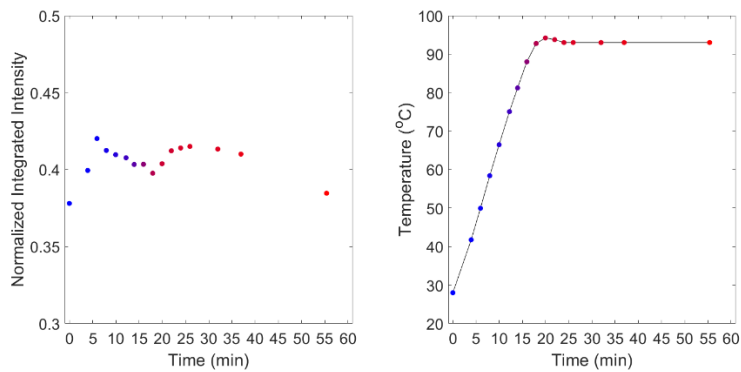

**Fig. S8G**

*Left* – Normalized integrated scattering intensity of the in situ reversion transformation at CHESS. There is a slight random deviation in the integrated intensity throughout the transformation. Including all data points, the deviation in intensity through the experiment is ~10%. By neglecting the two end points or pure phase F313 and F324, the average deviation becomes ~5%. This small deviation indicates no new species have been generated (from fragmentation) or lost (from melting). These new species would result in peak broadening and loss of scattering intensity. *Right* – The time-temperature profile of the in situ F313 to F324 transformation. This profile is used in the determination of the F313 fraction in **Fig. S8E**.

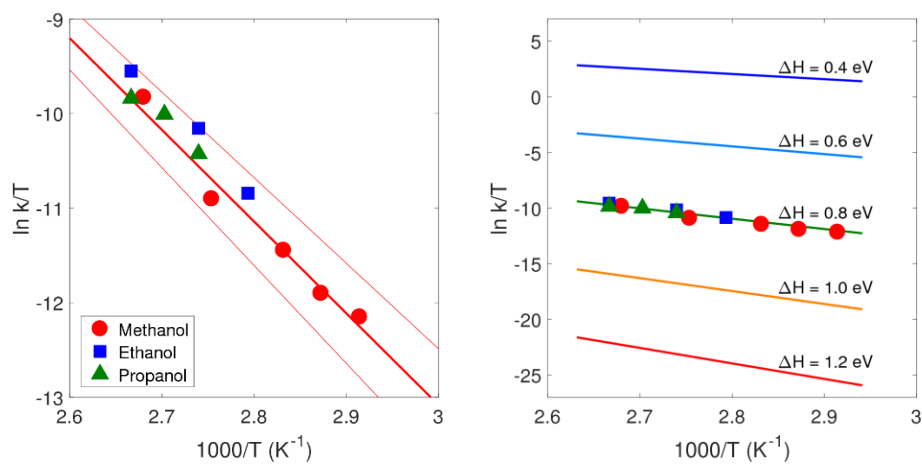

**Fig. S8H**

*Left* – Eyring plot of the reversion process for F313 produced from methanol (MeOH), ethanol (EtOH), and propanol (PrOH). The reversion rate constant using different alcohols are within the experimental uncertainty, which indicates the reversion process is independent of the alcohol. *Right* – The Eyring equation using different enthalpies (ΔH) with a constant entropy (ΔS = -0.75 meV/K). From theoretical calculations of adsorption/desorption of alcohols, as the chain length of the alcohol increases there is a change in the adsorption energy (~0.2 eV/carbon unit)(*44*). Presuming no configurational dependence (ΔS$_{configuration}$ = 0) of the alcohol chain length, the ΔS of the reversion process is constant and is a combination of vibrational and H-bonding entropy. Therefore, an increasing alcohol chain length would only affect ΔH if the energy barrier of the reversion process is from the adsorption/desorption process. Considering the rates of reversion are similar and there is no large deviation in the ΔH of the transformation, we conclude the apparent energy barriers are not associated with adsorption/desorption processes.

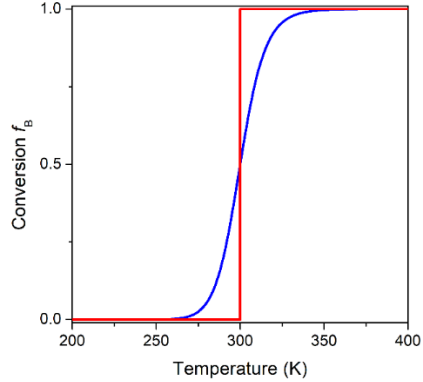

**Fig. S9**

Schematic of the dependence of the equilibrium extent of conversion to B, $f_B = [B]/([A]+[B])$, for the reaction A → B on the thermodynamic state variable (temperature) for a unimolecular reaction (blue) and a solid-solid phase transition (red). The degree of conversion depends exclusively on the change in Gibbs free energy of reaction $\Delta G_{A \to B} = \Delta H_{A \to B} + T\Delta S_{A \to B}$, where $\Delta H_{A \to B}$ is the enthalpy of reaction and $\Delta S_{A \to B}$ the entropy. In this example we assign the arbitrary values $\Delta H_{A \to B} = +1.0$ eV (~100 kJ/mol) and $\Delta S_{A \to B} = +3.5$ meV/K (~333.3 J/K·mol).

For the unimolecular reaction $f_B$ varies smoothly with T, and the degree of conversion is given by the equilibrium constant $K_{A \to B} = \exp[-\Delta G_{A \to B}/RT]$. The activities of A and B are related to $K_{A \to B} = [B]/[A]$, and $[A]+[B] = 1$, so that $f_B = [B] = K_{A \to B}/(1+K_{A \to B})$. Both A and B coexist at equilibrium across a wide range of T if they are described as isomers. Where A → B is described as a phase transition, coexistence of A and B is only allowed at exactly $T_c = 300$ K, where $\Delta G_{A \to B} = 0$. Above $T_c$, $\Delta G_{A \to B} < 0$ so that the system entirely consists of B; below $T_c$, $\Delta G_{A \to B} > 0$ and only A exists at equilibrium.

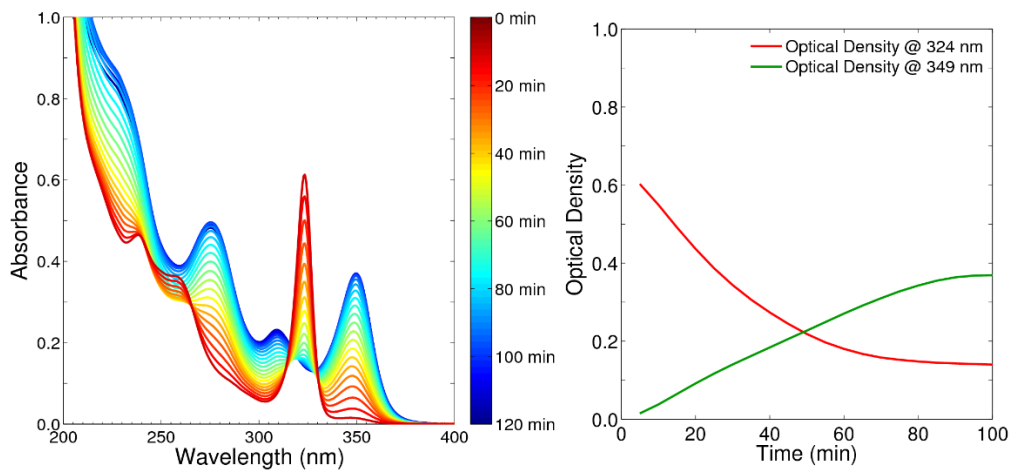

**Fig. S10**

*Left* – In-situ transformation of the F324 in hexane. *Right* – Evolution in the optical density of the spectra at 324 nm and 349 nm. 100 µL of ethanol (4.0 %vol.) is added to F324 dissolved in 2.5 mL of hexane. In a liquid cell, the F324 quickly diminishes with the introduction of ethanol at room temperature. However, rather than the formation of the F313, there is the discrete formation of a new peak at a wavelength of 349 nm. The rate of accumulation for the peak 349 nm is roughly the same as the rate of depletion of the optical density at 324 nm. After full consumption of the F324, the peak at 349 nm is roughly half the optical density of the initial F324, which suggests a perfect bimolecular collision of the F324 in a fluid medium. There are well resolved peaks of the new spectrum at 349 nm, 310 nm, and 275 nm. The 310 nm peak is suspected to be the F313 isomer and is shifted due to a dielectric solvatochromic shift (see **Fig. S6**). When the hexane and ethanol are evaporated the solid mass is re-dissolved in fresh hexane, the spectrum is that of a pristine F313 (see **Fig. S1**). We do not believe the formation of the spectrum with the peak at 349 nm to be another cluster or larger NP's, but a coupling effect between the hexane/alcohol mixture and the clusters. Further investigation is necessary.

**Table S1.**

Peak positions, standard deviations (Std. Dev.) and spectral areas of the cadmium $3d_{5/2}$ XPS peaks fitted to Gaussian functions. All values are extracted from the data in **Fig. S4A**. The intensity scale is normalized to 1 at the maximum of the peak. The spectral area is the sum of normalized intensity between the binding energies of 402 to 408 eV.

| | Cadmium $3d_{5/2}$ peak | | |
|---|---|---|---|
| **Sample** | **Peak (eV)** | **Std. Dev. (eV)** | **Spectral Area** |
| CdS Bulk | 405.1 | 0.54 | 23.5 |
| CdS NPs | 405.1 | 0.69 | 28.3 |
| Cd Oleate | 405.5 | 0.59 | 24.8 |
| F313 | 405.5 | 0.58 | 24.3 |
| F324 | 405.5 | 0.64 | 26.2 |

**Table S2.**

Peak positions, standard deviations (Std. Dev.) and spectral areas of the oxygen 1s XPS peaks fitted to Gaussian functions. All values are extracted from fits to the data in **Fig. S4A**. The intensity scale is normalized to 1 at the maximum of the C-O bonding peak. The experimental maxima (observed) peak is identified by **Peak – Obs**. For the C-O bonding peak, some samples have a shift in the peak position between the observed and fitted. The standard deviation within a sample is held constant between all of the fitted peaks.

| | Oxygen 1s – C-O Bonding | | | |
|---|---|---|---|---|
| **Sample** | **Peak – Obs. (eV)** | **Peak – Fit (eV)** | **Std. Dev. (eV)** | **Spectral Area** |
| CdS Bulk | 531.9 | 531.8 | 0.75 | 17.3 |
| CdS NPs | 531.9 | 531.7 | 0.87 | 20.0 |
| Cd Oleate | 531.7 | 531.7 | 0.70 | 16.1 |
| F313 | 531.7 | 531.7 | 0.65 | 15.6 |
| F324 | 531.9 | 531.9 | 0.70 | 17.0 |

| | Oxygen 1s – C=O Bonding | | |
|---|---|---|---|
| **Sample** | **Peak (eV)** | **Std. Dev. (eV)** | **Spectral Area** |
| CdS Bulk | 533.6 | 0.75 | 2.6 |
| CdS NPs | 533.7 | 0.87 | 4.5 |
| Cd Oleate | 533.6 | 0.70 | 3.5 |
| F313 | 533.6 | 0.65 | 1.8 |
| F324 | 533.6 | 0.70 | 1.8 |

| | Oxygen 1s – C-OH Bonding | | |
|---|---|---|---|
| **Sample** | **Peak (eV)** | **Std. Dev. (eV)** | **Spectral Area** |
| CdS Bulk | N/A | N/A | N/A |
| CdS NPs | N/A | N/A | N/A |
| Cd Oleate | 535.0 | 0.70 | 3.5 |
| F313 | 535.0 | 0.65 | 1.8 |
| F324 | N/A | N/A | N/A |

**Table S3.**

Properties of the suspending and precipitating (cleaning) solvents for the MSCs.

| Solvent | Dielectric Constant | Reference | MSC Interaction |
|---|---|---|---|
| Hexane | 1.90 | (*45*) | Soluble |
| Tetrahydrofuran | 7.40 | (*46*) | Soluble |
| Fluorohexane | 1.69 | (*47*) | Precipitate |
| Ethyl Acetate | 6.02 | (*45*) | Precipitate |
| Fluoroethanol | 24.32 | (*48*) | Precipitate |
| Acetone | 27.70 | (*45*) | Precipitate |
| Methanol | 32.63 | (*45*) | Precipitate |
| Ethanol | 24.30 | (*45*) | Precipitate |
| Propanol | 20.10 | (*45*) | Precipitate |
| Octanol | 10.30 | (*45*) | Precipitate |

**Table S4.**

Rate constants from fits to the kinetics of conversion and reversion processes. Uncertainty is indicated by Δk. $R^2$ is the coefficient of determination. Uncertainties are determined by the difference between rate constants determined from fitting the raw data with an exponential or the natural log (linearized) of the data with a line. Minimization of the exponential fit to the raw data and the linear fit to the linearized data yield slightly different rate constants in the 52°C and 60°C rate constants of the conversion process because of a slight change in transformation rate. The rate constant extracted from the linearized data best fits long times (conversion >70%), whereas the exponential fit to the raw data best fits intermediate times (30% < conversion < 70%). The difference between the two rate constants were added to the uncertainty. For all other temperatures, the rate constants minimized via the raw data or the linearized data are the similar.

**Conversion (MeOH)**

| Temperature | k (s$^{-1}$) | Δk (s$^{-1}$) | $R^2$ |
|---|---|---|---|
| 28°C | 9.32x10$^{-5}$ | 5.60x10$^{-8}$ | 0.9955 |
| 36°C | 2.30x10$^{-4}$ | 1.49x10$^{-7}$ | 0.9995 |
| 40°C | 3.13x10$^{-4}$ | 3.13x10$^{-7}$ | 0.9985 |
| 42°C | 3.83x10$^{-4}$ | 5.45x10$^{-6}$ | 0.9975 |
| 47°C | 9.93x10$^{-4}$ | 5.59x10$^{-6}$ | 0.9988 |
| 52°C | 1.65x10$^{-3}$ | 6.65x10$^{-5}$ | 0.9767 |
| 60°C | 3.90x10$^{-3}$ | 5.89x10$^{-4}$ | 0.9264 |
| 70°C | 7.98x10$^{-3}$ | 1.17x10$^{-4}$ | 0.9925 |

Fitting Function: $f_{324}(t) = e^{-kt}$

**Reversion (MeOH)**

| Temperature | k (s$^{-1}$) | Δk (s$^{-1}$) | $R^2$ |
|---|---|---|---|
| 70°C | 1.82x10$^{-3}$ | 1.90x10$^{-5}$ | 0.9994 |
| 75°C | 2.37x10$^{-3}$ | 1.24x10$^{-4}$ | 0.9963 |
| 80°C | 3.78x10$^{-3}$ | 1.14x10$^{-4}$ | 0.9974 |
| 90°C | 6.70x10$^{-3}$ | 1.90x10$^{-4}$ | 0.9980 |
| 100°C | 2.02x10$^{-2}$ | 1.92x10$^{-3}$ | 0.9808 |

Fitting Function: $f_{324}(t) = 1 - e^{-kt}$

**Table S5.**

Activation energy and prefactor for the conversion and reversion processes using the Arrhenius equation. Errors for the activation and prefactor are indicated by a Δ. Errors have been given only from the data on the Arrhenius plot and have not propagated from the rate constant fits. The activation energies of the conversion and reversion processes are equivalent within the error of the fit.

| Process | $E_a$ (eV) | $\Delta E_a$ (eV) | A ($s^{-1}$) | lnA | ΔlnA | $R^2$ |
|---|---|---|---|---|---|---|
| Conversion (MeOH) | 0.99 | 0.04 | $2.91 \times 10^{12}$ | 28.7 | 1.5 | 0.9895 |
| Reversion (MeOH) | 0.87 | 0.07 | $8.82 \times 10^{9}$ | 22.9 | 2.4 | 0.9783 |

Arrhenius Equation: $$ln|k| = lnA - \frac{E_a}{k_b T}$$

**Table S6.**

Thermodynamic parameters for the conversion and reversion processes using the Eyring equation. The Gibbs free energy, ΔG, has a temperature dependence and is averaged across all ΔG for the conversion and reversion processes. The enthalpy (ΔH) and entropy (ΔS) are extracted from the slope and intercept, respectively. The subscript "err" denotes the error to ΔH and ΔS. Errors have been given only from the data on the Eyring plot and have not propagated from the rate constant fits.

| Process | $\Delta G_{avg}$ (eV) | $\Delta H$ (eV) | $\Delta H_{err}$ (eV) | $\Delta S$ (meV/K) | $\Delta S_{err}$ (meV/K) | $R^2$ |
|---|---|---|---|---|---|---|
| Conversion (MeOH) | 1.01 | 0.96 | 0.04 | -0.15 | 0.13 | 0.9889 |
| Reversion (MeOH) | 1.08 | 0.84 | 0.07 | -0.67 | 0.21 | 0.9768 |

Eyring Equation: $ln\left|\frac{k}{T}\right| = ln\left|\frac{k_b}{h}\right| + \frac{\Delta S}{k_b} - \frac{\Delta H}{k_b T}$